\documentclass[11pt]{article}

\usepackage{hyperref}
\usepackage{natbib}
\usepackage{color, latexsym, amsfonts, amssymb, amsthm, amscd, amsmath, graphicx}
\usepackage{mathrsfs, bm}
\usepackage{multirow,booktabs}

\newcommand{\be}{\begin{equation}}
\newcommand{\ee}{\end{equation}}
\newcommand{\bas}{\begin{eqnarray*}}
\newcommand{\eas}{\end{eqnarray*}}
\newcommand{\ba}{\begin{eqnarray}}
\newcommand{\ea}{\end{eqnarray}}
\newcommand{\bit}{\begin{itemize}}
\newcommand{\eit}{\end{itemize}}
\newcommand{\ben}{\begin{enumerate}}
\newcommand{\een}{\end{enumerate}}

\newcommand{\blue}{\color{black}}

\newtheorem{theorem}{Theorem}

\newtheorem{remark}{Remark}

\newtheorem{proposition}{Proposition}

\newtheorem{definition}{Definition}

\begin{document}
%\doublespacing

\date{}
\title{
Statistical inference for the two-sample problem under likelihood ratio ordering, with application to the ROC curve estimation}

\author{Dingding Hu$^1$, Meng Yuan$^1$, Tao Yu$^2$ and Pengfei Li$^1$}

\maketitle

{\small
\begin{center}
$^1$Department of Statistics and Actuarial Sciences,
University of Waterloo,
Waterloo, ON,  N2L 3G1, Canada\\
%(Email: \emph{pengfei.li@uwaterloo.ca})\\

$^2$Department of Statistics and Data Science,
National University of Singapore,  117546, Singapore\\
%(Email: \emph{stayt@nus.edu.sg})\\

\end{center}

\hrule

{\small
\begin{quotation}
\noindent
{\bf Abstract:} {\blue The receiver operating characteristic (ROC) curve is a powerful statistical tool and has been widely applied in medical research. In the ROC curve estimation, a commonly used assumption is that larger the biomarker value, greater severity the disease. In this paper, we mathematically interpret ``greater severity of the disease" as ``larger probability of being diseased". This in turn is equivalent to assume the likelihood ratio ordering of the biomarker between the diseased and healthy individuals.  
With this assumption,} we first propose a Bernstein polynomial method to model the distributions of both samples; we then estimate the distributions by the maximum empirical likelihood principle. The ROC curve estimate and the associated summary statistics are obtained subsequently. 
{\blue Theoretically, we establish the asymptotic consistency of our estimators. Via extensive numerical studies, we compare the performance of our method with competitive methods.} The application of our method is illustrated by a real-data example. 

\vspace{0.3cm}

\noindent
{\bf Keywords:}  Area under the ROC curve, Bernstein polynomials,
Likelihood ratio ordering, ROC curve, Youden index.
\end{quotation}
}
\hrule

\bigskip

\bigskip

\section{Introduction}

{\blue The ROC curve is a powerful statistical tool and has been widely applied in many scientific areas; it evaluates the diagnostic abilities of a binary classifier for varied discrimination thresholds. Consider a medical study, where a biomarker serves as a binary classifier for diagnosing the healthy/diseased status of each patient. The ROC curve plots the ``sensitivity" against ``one minus the specificity" at all the possible thresholds of the biomarker. Here, sensitivity/specificity refers to the probability of correctly identifying the diseased/healthy individuals. The ROC curve related inferences have been extensively investigated in the literature; we refer to \cite{pepe2003}, \cite{Zhou2002}, \cite{Qin2003}, and \cite{chen2016using} for a comprehensive review of the existing developments.

Practically, for an individual, we may assume that the larger biomarker value indicates greater possibility/severity of the disease; therefore, we can classify individuals as diseased if their corresponding biomarkers exceed a given cutoff point $x$, and vice versa; see \cite{Yin2014} for a discussion. Under this assumption, the sensitivity and specificity are $1-F_1(x)$ and $F_0(x)$, where $F_0(\cdot)$ and $F_1(\cdot)$ denote the cumulative distribution functions (cdfs) of the biomarkers for individuals that are healthy and diseased, respectively. The ROC curve is then given by 
\begin{equation*}
\label{def.roc}
ROC(s)=1-F_1\big(F_0^{-1}(1-s)\big),
\end{equation*}
for $s\in(0,1)$. 

In the literature, there are two common strategies for mathematically accommodating the assumption that larger biomarker associates with the greater severity of the disease. The first interprets it as the biomarker values of diseased individuals are stochastically greater than those of healthy ones; that is $F_1(x) \leq F_0(x)$ for every $x\in \mathbb{R}$ \citep{Wang2017}. The second  assumes that larger biomarker value corresponds to larger probability of getting the disease \citep{Yu2017}; with the assumption that ``the larger biomarker value indicates greater severity of the disease", this second strategy is equivalent to interpret ``greater severity of the disease" as ``larger probability of being diseased". More specifically, let $X$ and $D$ denote the biomarker and indicator of the membership of an individual: $D=1$ or $0$ respectively indicates that the individual is diseased or healthy. Using Bayes' formula, we have 
\begin{equation}
\label{def.posterior}
P(D=1|X=x)
=\frac{P(D=1)f_1(x)}{P(D=1)f_1(x)+P(D=0)f_0(x)};
\end{equation}
where $f_1(x)$ and $f_0(x)$ are the probability density functions (pdfs) of the biomarkers
in the diseased population and the healthy population, respectively. 
Hence, the second interpretation equivalently assumes that $f_1(x)/f_0(x)$ is a monotone function of $x$, 
which is known as the likelihood ratio ordering \citep{Yu2017, Dykstra1995}. We observe that with the likelihood ratio ordering, the derivative of $ROC(s)$,
\begin{eqnarray*}
\frac{d\{ROC(s)\}}{d s} = \frac{f_0\left(F_1^{-1}(1-s)\right)}{f_1\left(F_1^{-1}(1-s)\right)},
\end{eqnarray*}
is nonincreasing, which implies that $ROC(s)$ is concave for $s\in(0,1)$. 
We refer to \cite{gneiting2018receiver} for the importance of concavity in the interpretation and modelling of ROC curves. Furthermore, we observe that as an assumption, likelihood ratio ordering is also widely adopted in many important applications of the Neyman–Pearson lemma, such as the Karlin–Rubin theorem.
In this paper, we shall incorporate the likelihood ratio ordering assumption to  improve the estimation of ROC curve and its summary statistics.

There are three types of estimation methods for the ROC curves: the parametric, the semiparametric \citep{zhou2008semi},  and the nonparametric methods \citep{Zhou2002}. We observe that many of these estimation methods did not incorporate the likelihood ratio ordering constraint; ignoring this constraint may result in less accurate ROC estimation. As far as we are aware, they are two estimation methods in the literature that have accommodated this constraint: \cite{Dykstra1995} considered the estimation of $F_0$ and $F_1$ with a maximum nonparametric likelihood approach; \cite{Yu2017} proposed a smoothed likelihood approach to estimate $f_0$ and $f_1$. 
The former leads to a non-smooth estimate of $P(D=1|X=x)$ in \eqref{def.posterior}, 
and therefore the resulting estimate of the optimal cutoff point that maximizes the Youden index is inefficient. See Section 1 of the supplementary material for more details.
} 
The latter relies on a smoothing parameter the optimal choice of which is not easily obtained.

In this paper, we propose a Bernstein polynomial  method that accommodates the likelihood ratio ordering assumption, and use the empirical likelihood principle to estimate $F_0$, $F_1$, ROC curve, and two popular summary statistics:  the area under the curve (AUC) and the Youden index \citep{youden1950index}. With the help of the existing R function \texttt{glmnet} \citep{friedman2021}, we establish an algorithm to implement our proposed method; we integrate our algorithm as an R package \texttt{BPLR} available at ``\texttt{https://github.com/Dingding-Hu/BPLR-package}". 
{\blue 
We have also established the asymptotic consistency of our method. It is noteworthy that the resulting estimate of $P(D=1|X=x)$ from our method is smooth; through extensive numerical studies, we observe that our ROC curve, Youden index, and the associated optimal cutoff point estimates outperform those from the existing methods in most examples.}

The rest of the paper is organized as follows. 
Section \ref{estimation} proposes the Bernstein polynomial method with the empirical likelihood principle to establish the estimators for $F_0(\cdot)$ and $F_1(\cdot)$, 
and subsequently result in the estimators of the ROC curve and its summary statistics. 
{\blue The consistency of the proposed method is also studied.}
Section \ref{simulation} presents the simulation results.
Section \ref{realdata} compares our method with existing methods in a real-data example, and Section \ref{conclude} concludes the paper with some discussion. 
{\blue Technical details are given in the supplementary material. }

\section{Main results}
\label{estimation}

\subsection{Bernstein polynomials approach}
Denote by $\{X_1,\ldots,X_{n_0}\}$  and $\{Y_1,\ldots,Y_{n_1}\}$
the random samples of biomarkers from the healthy and the diseased populations, respectively. 
The pdf and cdf of $X_i$'s are $f_0$ and $F_0$;  
the pdf and cdf of $Y_j$'s are $f_1$ and $F_1$. 
Let $n=n_0+n_1$ be the total sample size; 
denote by $t_1<\ldots<t_m$ the distinct values of the combined sample. 

We assume that $f_0$ and $f_1$ satisfy the likelihood ratio ordering, i.e., 
$f_1
(x)/f_0(x)$ is an increasing function of $x$. 
To incorporate this constraint in the estimation procedure, 
we propose to model $\log \{f_1(x)/f_0(x)\}$ as a linear combination of Bernstein polynomials. 
The definition of the Bernstein polynomials is given below. 

\begin{definition}
(Bernstein polynomials) Let $N$ be a  positive integer. For each $0\leq l\leq N$, the Bernstein polynomials are defined to be
\begin{equation*}
B_{l}(x;N)={N\choose l}x^l(1-x)^{N-l},~~l=0,...,N,~~x\in [0,1].
\end{equation*}
\end{definition}

Bernstein polynomials can serve as a set of base functions in the expansion of a nonparametric component. A nice feature of such an expansion is that it is able to incorporate the shape constraints as some condition(s) of the coefficients of polynomials. For a monotonic nonparametric component, we can expand it to be 
$\sum_{l=0}^N \beta_l B_{l}(x;N)$, which is an increasing function of $x$ if 
\begin{equation}
\label{bp.const}
\beta_0\leq\beta_1\ldots\leq \beta_N. 
\end{equation}
We refer to  \cite{WANG20122729}  for more details of incorporating other types of shape constraints in the Bernstein polynomial expansion. 

Next, we apply the Bernstein polynomials to our estimation problem. For presentational convenience, in the development below, we assume that $N$ is given and all biomarkers are ranged in $[0,1]$. Remarks \ref{remark-1} and \ref{remark-2} give the solutions of how to choose $N$ and transform the biomarkers so that they are contained in $[0,1]$ in practice. Based on the likelihood ratio ordering assumption, $\log\{f_1(x)/f_0(x) \} $ is an increasing  function of $x$. 
We propose to model it by a linear combination of Bernstein polynomials: 
\begin{equation}
\label{model.theta.bp0}
\log\{f_1(x)/f_0(x) \} = \sum_{l=0}^N \beta_l B_{l}(x;N),
\end{equation}
with $(\beta_0,\ldots,\beta_N)$ satisfying \eqref{bp.const}; and then 
incorporate the maximum empirical likelihood to estimate $(\beta_0,\ldots,\beta_N)$, $F_0$, and $F_1$. To this end,
let $$
a_i=\sum_{j=1}^{n_0}I(X_j=t_i)~~\mbox{ and }~~
b_i=\sum_{j=1}^{n_1}I(Y_j=t_i).
$$
Based on the observed two-sample data, 
the full likelihood function is given by
\begin{equation*}
\label{lik0}
L = \prod_{i=1}^m \{f_0(t_i)^{a_i}\cdot \{f_1(t_i)\}^{b_i}.
\end{equation*}
For $i=1,\ldots,m$, let
$$
p_{i0}=f_0(t_i)
~~\mbox{ and }~~
p_{i1}=f_1(t_i).
$$ 
Then, the empirical likelihood \citep{owen2000} is 
\begin{equation*}
\label{el.lik0}
L = \prod_{i=1}^m p_{i0}^{a_i}\cdot p_{i1}^{b_i}.
\end{equation*}

{\blue 
We assume $n_1/n \to \lambda \in (0, 1)$ as $n\to \infty$. For simplicity, hereafter we write $\lambda = n_1/n$ and assume that it is constant, since it does not affect our technical development.}
We further define
$$
\phi_i=(1-\lambda) f_0(t_i)+\lambda f_1(t_i)=(1-\lambda) p_{i0}+\lambda p_{i1}.
$$ 
By model \eqref{model.theta.bp0}, we have
\begin{eqnarray*}
\hspace{-0.2in}
\label{chapter3.relation1}
p_{i0}&=&\frac{\phi_i}
{1-\lambda+\lambda \exp\left\{ \sum_{l=0}^N \beta_l B_{l}(t_i;N)\right\} },\\
p_{i1}&=&\frac{\exp\left\{ \sum_{l=0}^N \beta_l B_{l}(t_i;N)\right\} \phi_i}
{1-\lambda+\lambda \exp\left\{ \sum_{l=0}^N \beta_l B_{l}(t_i;N)\right\} }.\label{chapter3.relation2}
\end{eqnarray*}
That is, $p_{i0}$'s and $p_{i1}$'s are determined by $\phi_i$'s and $(\beta_0,\ldots,\beta_N)$. 

With the above reparameterization, the empirical likelihood function of $(\phi_1,\ldots,\phi_m,\beta_0,\ldots,\beta_N)$ is then given as 
 \begin{equation*}
\label{chapter3.lik}
L=\prod_{i=1}^m   {p_{i0}}^{a_i} {p_{i1}}^{b_i} 
=(\lambda)^{-n_1}(1-\lambda)^{-n_0}\cdot 
L_1(\phi_1,\ldots,\phi_m)\cdot L_2(\beta_0,\ldots,\beta_N), 
\end{equation*}
where 
\begin{equation*}
\label{chapter3.lik12}
L_1(\phi_1,\ldots,\phi_m)=\prod_{i=1}^m \phi_i^{a_i+b_i}~~\mbox{ and }~~
L_2(\beta_0,\ldots,\beta_N)=  \prod_{i=1}^m \left[ \left\{ \theta(t_i) \right\}^{b_i}\{ 1-\theta(t_i)\} ^{a_i}\right]
\end{equation*}
with 
\begin{equation*}
%\label{model.theta.bp}
\theta(x)=\frac{\lambda \exp\left\{ \sum_{l=0}^N \beta_l B_{l}(x;N)\right\} }
{1-\lambda+\lambda \exp\left\{ \sum_{l=0}^N \beta_l B_{l}(x;N)\right\} }.
\end{equation*}
Note that feasible $\phi_i$'s satisfy 
\begin{equation}
\label{chapter3.cdf1}
\phi_i\geq 0,~~\sum_{i=1}^m\phi_i=1
\end{equation}
and 
\begin{equation}
\label{chapter3.cdf2}
\sum_{i=1}^m\phi_i \theta(t_i) =\lambda
\end{equation}
to  ensure that both $F_0$ and $F_1$ are cdfs. 
The  maximum empirical likelihood estimator (MELE) of $(\phi_1,\ldots,\phi_m,\beta_0,\ldots,\beta_N)$ is then defined to be
$$
(\hat\phi_1,\ldots,\hat\phi_m,\hat\beta_0,\ldots,\hat\beta_N)
=\arg\max_{\phi_1,\ldots,\phi_m,\beta_0,\ldots,\beta_N}L
$$
subject to constraints \eqref{chapter3.cdf1}, \eqref{chapter3.cdf2}, and the inequality constraints in \eqref{bp.const}.  

To solve the optimization problem above, we consider the following reparameterization of  $\beta_l$'s: 
$$
\beta_0=\alpha_0,~~\beta_1=\alpha_0+\alpha_1,\ldots,\beta_N=\sum_{l=0}^N \alpha_l.
$$ 
Then \eqref{bp.const} is equivalent to 
$$
\alpha_1\geq0,\ldots,\alpha_N\geq0.
$$
The above reparameterization implies that
\begin{equation*}
\sum_{l=0}^N \beta_l B_{l}(x;N)
=\sum_{l=0}^N \alpha_l B_{l}^*(x;N),
\end{equation*}
where $B_l^*(x;N)=\sum_{k=l}^N B_k(x;N)$  for $l=1,\ldots,N$ and $B_0^*(x;N)=1$. 

With a slight abuse of notation, we write 
\begin{equation*}
L_2(\alpha_0,\ldots,\alpha_N)=  \prod_{i=1}^m \left[ \left\{ \theta(t_i) \right\}^{b_i}\left\{ 1-\theta(t_i)\right\} ^{a_i}\right]
\end{equation*}
with 
\begin{equation*}
%\label{model.theta.bp}
\theta(x)=\frac{\lambda \exp\left\{\alpha_0+ \sum_{l=1}^N \alpha_l B_{l}^*(x;N)\right\} }
{1-\lambda+\lambda \exp\left\{\alpha_0+ \sum_{l=1}^N \alpha_l B_{l}^*(x;N) \right\} }.
\end{equation*}

The following proposition summarizes the results for calculating the MELEs of $\phi_i$'s and 
$(\alpha_0,\alpha_1,\ldots,\alpha_N)$. The proof is given in the supplementary material.

\begin{proposition}
\label{prop1}
Let 
$$
(\hat \phi_1,\ldots,\hat\phi_m)=\arg\max_{\phi_1,\ldots,\phi_m}
L_1(\phi_1,\ldots,\phi_m) \mbox{ subject to }\eqref{chapter3.cdf1}
$$
and 
$$
(\hat \alpha_0,\ldots,\hat\alpha_N)
=\arg\max_{\alpha_0,\ldots,\alpha_N} L_2(\alpha_0,\ldots,\alpha_N)
$$ 
subject to $\alpha_l\geq0$ for $l=1,\ldots,N$. 
Then 
\begin{enumerate}
\item[(a)] $\hat\phi_i=(a_i+b_i)/n$ for $i=1,\ldots,m$; 
\item[(b)]  $\sum_{i=1}^m\hat\phi_i  \hat \theta(t_i)=\lambda $, 
where 
\begin{eqnarray}
\hat\theta(x)
=\frac{\lambda \exp\left\{\hat \alpha_0+ \sum_{l=1}^N \hat\alpha_l B_{l}^*(x;N)\right\} }
{1-\lambda+\lambda \exp\left\{\hat \alpha_0+ \sum_{l=1}^N\hat \alpha_l B_{l}^*(x;N) \right\} }. \label{eq-theta-hat}
\end{eqnarray}
\end{enumerate}
\end{proposition}

 Proposition \ref{prop1} implies that we can maximize $L_1$ and $L_2$ separately to obtain the MELEs of $\phi_i$'s and 
$(\alpha_0,\alpha_1,\ldots,\alpha_N)$. 
The MELEs of $\phi_i$'s have the closed form in  Proposition \ref{prop1} (a). 
 Note that $L_2(\alpha_0,\ldots,\alpha_N)$ can be viewed as the likelihood for the standard logistic regression with the intercept being $\alpha_0+\log\{\lambda/(1-\lambda)\} $
 and covariates being $B_{1}^*(x;N),\ldots, B_{N}^*(x;N)$. 
 Then 
   $\left(\hat \alpha_0,\ldots,\hat\alpha_N\right)$  
   can be readily calculated by using the existing R function \texttt{glmnet}.
 Once $\hat\phi_i$'s and  $\left(\hat \alpha_0,\ldots,\hat\alpha_N\right)$   are available, 
the estimates of $p_{i0}$'s and $p_{i1}$'s are given by
\begin{eqnarray*}
\hat p_{i0}&=&\frac{\hat\phi_i}
{ 1-\lambda+\lambda \exp\left\{\hat \alpha_0+ \sum_{l=1}^N\hat \alpha_l B_{l}^*(t_i;N) \right\}  }\\
\hat p_{i1}&=&\frac{ \exp\left\{\hat \alpha_0+ \sum_{l=1}^N\hat \alpha_l B_{l}^*(t_i;N) \right\} \hat \phi_i}
{1-\lambda+\lambda \exp\left\{\hat \alpha_0+ \sum_{l=1}^N\hat \alpha_l B_{l}^*(t_i;N) \right\}   },
\end{eqnarray*}
which lead to the estimates for $F_0(\cdot)$ and $F_1(\cdot)$:
\begin{equation}
\label{hatF.def}
\hat F_{0}(x)=\sum_{i=1}^m \hat  p_{i0} I(t_i\leq x) ~~\mbox{ and }~~\hat F_{1}(x)=\sum_{i=1}^m \hat p_{i1} I( t_i\leq x).
\end{equation}

We make some remarks for the proposed method above. 
\begin{remark} \label{remark-1}
In the development above, we have assumed that $N$ is known. In practice, we can choose it based on the Bayesian information criterion (BIC). 
{\blue 
Specifically, since we maximize $L_2(\alpha_0,\ldots,\alpha_N)$ to obtain the estimators of $(\alpha_0,\ldots,\alpha_N)$, 
we can use it to establish the BIC criterion. 
For any $N$, let 
$$
(\tilde \alpha_0,\tilde\alpha_1,\ldots,\tilde\alpha_N)
=\arg\max_{\alpha_0,\ldots,\alpha_N} L_2(\alpha_0,\ldots,\alpha_N) 
$$
and 
$$
\tilde\theta(x)
=\frac{\lambda \exp\left\{\tilde \alpha_0+ \sum_{l=1}^N \tilde\alpha_l B_{l}^*(x;N)\right\} }
{1-\lambda+\lambda \exp\left\{\hat \alpha_0+ \sum_{l=1}^N\tilde \alpha_l B_{l}^*(x;N) \right\} }.
$$
That is, $\tilde\theta(x)$ is the MELE of $\theta(x)$ without the monotonicity assumption. 
Denote by $df_N$
the number of unknown parameters in $\theta(x)$. 
We define 
$$
BIC(N)=-2\log\left[ \prod_{i=1}^m  \{  \tilde\theta_N(t_i)\} ^{b_i}\{ 1-\tilde \theta_N(t_i)\}^{a_i}\right]+\left(\log n\right) \cdot   df_N. 
$$
Consequently, $N$ is set to be the minimizer of $BIC(N)$.
We use $\tilde\theta(x)$ instead of $\hat\theta(x)$
to construct the BIC criterion because without the monotonicity assumption, the number of unknown parameters in $\theta(x)$ can be clearly counted. 
}
\end{remark}

\begin{remark} \label{remark-2}

Practically, a biomarker, $x$ say, may not be in the range $[0,1]$. We can consider the transformation 
$$
x^*=\frac{x-t_{(1)}}{t_{(m)}-t_{(1)}},
$$
with $t_{(1)}$ and $t_{(m)}$ being the minimum and the maximum values of $t_i$'s. Clearly $x^*\in[0,1]$; we can then apply our method to the transformed biomarkers. 

\end{remark}

\begin{remark} \label{remark-3}

We observe that biomarkers may exhibit high variability in practice; applying a log transformation on them may improve the performance. Furthermore, we may include both the original and transformed biomarkers in the model. Specifically, we may consider  
\begin{equation}
\label{lr.bp.both}
\log\{f_1(x)/f_0(x)\}
=\alpha_0+\sum_{l=1}^N\alpha_l B_{l}^*(x^*;N)+\sum_{l=1}^N\alpha_{N+l} B_{l}^*(z^*;N),
\end{equation}
where 
$$
z^*=\frac{\log x-\log t_{(1)}}{\log t_{(m)}-\log t_{(1)}},
$$
and $\alpha_l\geq 0$ for $l=1,\ldots,(2N)$. 
All our developments above can be similarly applied to \eqref{lr.bp.both}.
This modelling strategy is applied in all the simulation and real data examples.

\end{remark}

\begin{remark}
\label{remark-4}
{\blue Our method is established on the likelihood ratio ordering assumption. To check the rationale of this assumption in practice, we can plot  $(\hat F_0,\hat F_1)$ in \eqref{hatF.def} and $(\tilde F_0,\tilde F_1)$, where $\tilde F_0$ and $\tilde F_1$
are respectively the empirical cdfs of $\{X_1,\ldots,X_{n_0}\}$  and $\{Y_1,\ldots,Y_{n_1}\}$. 
If the plots of $\hat F_i$ are reasonably close to those of $\tilde F_i$ for $i=0, 1$, we may regard the likelihood ratio ordering as a reasonable assumption. More rigorously, we can also use the goodness-of-fit test statistics:
$$
\Delta_n=\sup_{x}|\hat F_0(x)-\tilde F_0(x)|,
$$
and the Bootstrap method to test this assumption. 
}

\end{remark}

\subsection{Estimation of the ROC curve and its summary statistics}
\label{est.summary}

With the estimate $\hat F_0(x)$ and $\hat F_1(x)$, the ROC curve can be estimated by 
\begin{eqnarray}
\label{hatroc.def}
\widehat {ROC} (s)&=&1-\hat F_1\left(\hat F_0^{-1}(1-s)\right). 
\end{eqnarray}
In this paper, we further consider two summary statistics based on the ROC curve: AUC and Youden index. 
The AUC is the total area under the ROC curve, i.e., 
\begin{equation*}
\label{def.auc}
AUC=\int_0^1 ROC(s)~ds.
\end{equation*}
Note that for a given threshold or cutoff point of the biomarker, it is desirable to have
the corresponding value on $y$-axis (the sensitivity) to be as large as possible. 
Hence, the larger AUC value indicates that the binary classifier has stronger classification ability. With $\widehat {ROC} (s)$, AUC can be estimated by 
\begin{eqnarray}
\label{hatauc.def}
\widehat {AUC} &=&\int_{0}^1 \widehat {ROC} (s) ds.
\end{eqnarray}
The Youden index ($J$) is defined as the maximum value of  the sensitivity plus the specificity minus 1, 
i.e.,  
\begin{equation*}
\label{def.youden}
J=\max_{x}\{1-F_1(x)+F_0(x)-1\}=\max_{x}\{F_0(x)-F_1(x)\}. 
\end{equation*}
One advantage of Youden index is that it results in a criterion to choose the ``optimal" cutoff point, 
 which is the ``$x$" where Youden index is achieved. Specifically
  \begin{equation}
  \label{def.cutoff}
C=\arg\max_x \{F_0(x)-F_1(x)\}. 
\end{equation}
We refer to \cite{Yuan2021} and the references therein for recent developments in the Youden index and the optimal cutoff point estimation.
Note that \eqref{def.cutoff} implies $f_0(C)=f_1(C)$, and hence $\theta(C)=\lambda$. Therefore we can solve the equation 
\begin{equation}
\label{hatC.def}
\hat\theta(\hat C)=\lambda
\end{equation}
to obtain $\hat C$
and as a consequence, 
\begin{equation}
\label{hatJ.def}
\hat J= \hat F_{0}(\hat C)-\hat F_{1}(\hat C).
\end{equation}

  {\blue 
 \subsection{Asymptotic properties}
 In this section, we establish the consistency of our estimators. We need the following notation. 
Let $G(x)=\lambda F_1(x)+(1-\lambda)F_0(x)$ and denote
\begin{equation}
\label{def.theta0}
\theta_0(x)=\frac{\lambda f_1(x)}{(1-\lambda)f_0(x)+ \lambda f_1(x)}.
\end{equation}
We first show the $L_2$ convergence of $\hat \theta(\cdot)$ to $\theta_0(\cdot)$, where $\hat \theta(\cdot)$ is defined by \eqref{eq-theta-hat}. For presentational continuity, we give the conditions in the Appendix, and relegate the technical details to the supplementary material.

 \begin{theorem}
 \label{thm1}
 Assume Conditions A1--A3 in the Appendix. 
We have 
\begin{eqnarray*}
\int_{0}^1 \left\{\hat\theta(x)- \theta_0(x)\right\}^2~dG(x)
=o_p(1). 
\end{eqnarray*}
 \end{theorem}

With Theorem \ref{thm1}, 
  we are able to establish the asymptotic consistency of 
 $(\hat F_0,\hat F_1)$ in \eqref{hatF.def}, $\widehat {ROC} (s)$ in \eqref{hatroc.def},
 $\widehat {AUC} $ in \eqref{hatauc.def}, 
$\hat C$ in \eqref{hatC.def}, and $\hat J$ in \eqref{hatJ.def}; the results are given in Theorem \ref{theorem-2} below.

 \begin{theorem} \label{theorem-2}
Assume Conditions A1--A3 in the Appendix. 
We have 
\begin{enumerate}
    \item[(a)] $\sup_{x\in [0,1]} \left|\hat F_0(x)-F_0(x)\right|=o_p(1)~\text{and}~\sup_{x\in [0,1]} \left|\hat F_1(x)-F_1(x)\right|=o_p(1)$,
    \item[(b)] $\sup_{s\in [0,1]} \left|\widehat{ROC}(s)-ROC(s)\right|=o_p(1)$,
    \item[(c)] $\widehat{AUC}=AUC+o_p(1)$.
    \end{enumerate}
    Furthermore, if Condition A4 in the Appendix is also satisfied, then 
\begin{enumerate}
    \item[(d)] $\hat{C}=C+o_p(1)$ \quad and \quad $\hat{J}=J+o_p(1)$.
\end{enumerate}
\end{theorem}
}

\section{Simulation study}
\label{simulation}

 \subsection{Simulation setup}
In this section, with simulation examples, we compare the performance of our proposed method (denoted as ``BP") 
with existing methods in the estimation of the ROC curves and its summary statistics. We consider the following {\blue seven} competitive methods: 
 \begin{enumerate}
 \item[--]  the Box-Cox method in \cite{bantis_nakas_reiser_2018}, denoted as ``Box-Cox";
 {\blue \item[--] the method in \cite{zhou2008semi} under the binormal model, denoted as ``ZL";}
 {\blue \item[--] the method in \cite{lin2012direct}, denoted as ``LZL";}
 \item[--]  the ECDF-based method, denoted as ``ECDF", {\blue which is also known as the nonparametric estimator in  \cite{pepe2003} and \cite{Zhou2002}}; 
 \item[--] the maximum nonparametric likelihood method under the likelihood ratio ordering \citep{Dykstra1995}, denoted as ``MNLE";
 \item[--]  the kernel-based method in \cite{bantis_nakas_reiser_2018}, denoted as ``Kernel";
  \item[--] the maximum smoothed likelihood method under the likelihood ratio ordering in \cite{Yu2017}, denoted as ``MSLE". 
 \end{enumerate}
 
 In the Kernel and MSLE methods, we follow \cite{bantis_nakas_reiser_2018} and \cite{Yu2017}.
 We use the Gaussian kernel with bandwidths $h_0$ and $h_1$ for healthy and diseased groups respectively, where $$
h_0=0.9\min\left\{s_0,\frac{q_0}{1.34}\right\}n_0^{-0.2}
~~\mbox{ and }~~
h_1=0.9\min\left\{s_1,\frac{q_1}{1.34}\right\}n_1^{-0.2}.
$$
Here, $s_0$ and $q_0$ are sample standard deviation and sample interquartile range for the sample from the healthy population,
and $s_1$ and $q_1$ are sample standard deviation and sample interquartile range for the sample from  the diseased population. 

{\blue Throughout our numerical studies, we observe that the MNLE and MSLE methods have similar performance as that of the ECDF and kernel methods, respectively. Furthermore, the LZL method is designed to accommodate covariates; without covariates, the performance of this method is very similar to that of the ECDF method. For space limitation, we present the results for MNLE, MSLE, and LZL methods in the supplementary material. }

We consider two distributional settings: 
 \begin{enumerate}
 \item[(1)] $f_0\sim N(\mu_0,\sigma_0^2)$ and $f_1\sim N(\mu_1,\sigma_1^2)$;
 \item[(2)] {\blue $f_0\sim Gamma(a_0,b_0)$ and $f_1\sim Gamma(a_1,b_1)$.} 
 \end{enumerate}
 {\blue Here, $Gamma(a,b)$ denotes the gamma distribution with shape parameter $a$ and rate parameter $b$.}
 Note that Setting (1) corresponds to the case that the model assumption for the Box-Cox method is satisfied, whereas Setting (2) corresponds to the case that the model assumption for the Box-Cox method is violated. 
In both cases,  $f_0$ is fixed. 
The parameters for $f_1$ are varied such that the corresponding Youden indices are 0.3, 0.5, and 0.7, respectively.
The details of these settings are given in Table \ref{table:nonlin}.
{\blue We consider three different combinations of sample sizes: $(n_0,n_1)=(50,50), (100,100)$, and $(150,50)$.
Therefore, for each distributional setting of $f_0$ and $f_1$, we have 9 combinations of parameters and sample sizes. For each combination, we repeat the simulation 2000 times. 
}

\begin{table}[!http]
\caption{Simulation settings}
\centering
\begin{tabular*}{\textwidth}{c@{\extracolsep{\fill}} c c cc  c c} 
\toprule
Distribution & $J$ & $AUC$ & $\mu_0$ & $\sigma_0^2$& $\mu_1$ & $\sigma_1^2$  \\
\midrule
Normal  &0.3 & 0.707 & 10 & 1 & 10.771 & 1\\
Normal  &0.5 & 0.830 & 10 & 1 & 11.349 & 1\\
Normal  &0.7 & 0.929 & 10 & 1 & 12.073 & 1\\
\midrule
Distribution & $J$ & $AUC$ & $a_0$ & $b_0$& $a_1$ & $b_1$  \\
\midrule
%Beta &0.3 & 0.702 & 2 & 2 & 3.838 & 2\\
%Beta &0.5 & 0.832 & 2 & 2 & 4 & 1.166\\
%Beta &0.7 & 0.931 & 2 & 2 & 5 & 0.676\\
%\midrule
%Distribution & $J$ & $AUC$ & $a_0$ & $b_0$& $a_1$ & $b_1$  \\
%\midrule
Gamma &0.3 & 0.708 & 2 & 1 & 3 & 0.937\\
Gamma &0.5 & 0.830 & 2 & 1 & 4 & 0.944\\
Gamma &0.7 & 0.929 & 2 & 1 & 5 & 0.827\\
\bottomrule

\end{tabular*}
\label{table:nonlin}
\end{table}

{\blue We observe that the proposed method depends on the choice of $N$. In the supplementary material, we give the frequency of the selected $N$ values based on the BIC criterion given in Remark \ref{remark-1} from 2000 repetitions. Furthermore, we have considered the distributional setting that $f_0$ and $f_1$ follow the Beta distributions; the details and results are also given in the supplementary material.
} 

\subsection{Comparison of the ROC curve estimation}

In this section, we compare our proposed method with the competitive methods in the ROC curve estimation. 
The criteria for comparison are the $L_1$- and $L_2$-distances between the estimated and the true ROC curves.
For a generic ROC curve estimate $\overline {ROC}(s)$, the $L_1$- and $L_2$-distances between $\overline {ROC}(s)$ and  the true ROC curve $ROC(s)$ are defined to be:
\begin{equation*}
L_1(\overline {ROC},ROC)= \int_{0}^1 \left|\overline {ROC}(s)-ROC(s)\right|~ds
\end{equation*}
and 
\begin{equation*}
L_2(\overline {ROC},ROC)=\left[ \int_{0}^1 \left\{ \overline{ROC}(s)-ROC(s)\right\}^2~ds\right]^{1/2}.
\end{equation*}
Simulation results based on 2000 repetitions are summarized in Table \ref{table:roc}.

{\blue From this table and the results given in the supplementary material, we observe that our BP method results in the smallest average $L_1$- and $L_2$-distances for all the examples; the improvement is significant. For example,  the ratio of the $L_2$-distances between our BP method and the ECDF method is ranged between 0.61 and 0.71 with an average of 0.66; the ratio of the $L_2$-distances between our BP method and the ZL method is between 0.84 and 0.93 with an average of 0.88. }

\begin{table}[!htbp]
\caption{Averages of $L_1$-distances and $L_2$-distances of five methods for estimating the ROC curve}
\centering
%\small
\begin{tabular*}{\textwidth}{c@{\extracolsep{\fill}} c c c c c c c} 
\toprule
 &$(n_0,n_1)$ & \multicolumn{2}{c}{$(50,50)$} &  \multicolumn{2}{c}{$(100,100)$} &  \multicolumn{2}{c}{$(150,50)$} \\ 
 \cline{3-8}

Distribution ($J$)&Method & $L_1$ & $L_2$ & $L_1$& $L_2$ & $L_1$ & $L_2$ \\
\midrule
Normal (0.3)& BP&0.040&0.046&0.029&0.033&0.033&0.038\\
&Box-Cox & 0.044 & 0.053 & 0.032 &  0.038 & 0.037 & 0.043\\
&ZL & 0.045 & 0.054 & 0.032 & 0.038 & 0.037 & 0.044\\
% &LZL&0.056&0.071&0.040&0.051&0.046&0.058\\
& ECDF & 0.056 & 0.071 & 0.040 & 0.051& 0.046 & 0.058 \\
%& MNLE & 0.054 & 0.066 & 0.038 & 0.047 & 0.045 & 0.054 \\
& Kernel & 0.047 & 0.057 & 0.035 & 0.042 & 0.040 & 0.048 \\
%& MSLE & 0.045 & 0.054 & 0.035 & 0.041 & 0.039 & 0.046\\
\midrule
Normal (0.5) &BP&0.032&0.041&0.023&0.029&0.026&0.034\\
&Box-Cox & 0.036 & 0.047 & 0.025 &  0.034 & 0.029 & 0.038\\
&ZL & 0.036 & 0.049 & 0.026 & 0.034 & 0.030 & 0.039\\
% &LZL&0.045&0.064&0.032&0.046&0.037&0.051\\
& ECDF & 0.045 & 0.064 & 0.032 & 0.046 & 0.037 & 0.051\\
%& MNLE & 0.042 & 0.059 & 0.030 & 0.043 & 0.035 & 0.047\\
& Kernel & 0.040 & 0.053 & 0.029 & 0.040 & 0.033 & 0.043 \\
%& MSLE & 0.039 & 0.052 & 0.029 & 0.039 & 0.033 & 0.043\\
\midrule
Normal (0.7) &BP&0.020&0.033&0.014&0.023&0.016&0.026\\
&Box-Cox & 0.022 & 0.036 & 0.015 &  0.026 & 0.017 & 0.028\\
&ZL & 0.023 & 0.039 & 0.016 & 0.027 & 0.019 & 0.030\\
%&LZL&0.029&0.053&0.020&0.038&0.023&0.040\\
& ECDF & 0.029 & 0.053 & 0.020 & 0.038 & 0.023 & 0.040\\
%& MNLE & 0.026 & 0.047 & 0.019 & 0.035 & 0.021 & 0.036 \\
& Kernel & 0.027 & 0.044 & 0.020 & 0.033 & 0.022 & 0.035\\
%& MSLE & 0.027 & 0.044 & 0.020 & 0.033 & 0.022 & 0.035\\
\midrule
Gamma (0.3)& BP&0.042&0.050&0.030&0.035&0.035&0.041\\
&Box-Cox&0.045&0.054&0.032&0.038&0.037&0.044\\
&ZL&0.047&0.056&0.032&0.039&0.038&0.045\\
%&LZL&0.057&0.072&0.040&0.052&0.046&0.058\\
&ECDF&0.057&0.072&0.040&0.052&0.046&0.058\\
%&MNLE&0.054&0.067&0.039&0.048&0.044&0.054\\
&Kernel&0.051&0.062&0.038&0.046&0.044&0.051\\
%&MSLE&0.049&0.059&0.037&0.044&0.042&0.049\\
\midrule
Gamma (0.5) &BP&0.034&0.046&0.024&0.032&0.027&0.035\\
&Box-Cox&0.036&0.049&0.026&0.035&0.028&0.037\\
&ZL&0.037&0.050&0.026&0.035&0.028&0.038\\
%&LZL&0.045&0.067&0.032&0.047&0.036&0.050\\
&ECDF&0.045&0.067&0.032&0.047&0.036&0.050\\
%&MNLE&0.042&0.061&0.031&0.044&0.034&0.047\\
&Kernel&0.046&0.060&0.035&0.044&0.038&0.046\\
%&MSLE&0.045&0.058&0.034&0.044&0.037&0.046\\
\midrule
Gamma (0.7) &BP&0.021&0.036&0.016&0.026&0.016&0.027\\
&Box-Cox&0.022&0.038&0.016&0.027&0.016&0.028\\
&ZL&0.023&0.040&0.016&0.028&0.018&0.031\\
%&LZL&0.029&0.055&0.021&0.040&0.022&0.040\\
&ECDF&0.029&0.055&0.021&0.040&0.022&0.040\\
%&MNLE&0.025&0.049&0.019&0.036&0.020&0.036\\
&Kernel&0.033&0.050&0.026&0.038&0.028&0.038\\
%&MSLE&0.033&0.049&0.025&0.037&0.027&0.038\\
\bottomrule
\end{tabular*}
\label{table:roc}
\end{table}

\subsection{Comparison of the AUC estimation}
\label{simu.auc}
In this section, we compare our proposed method with the competitive methods in the AUC estimation. 
The criteria for comparison are the relative bias (RB) and mean square error (MSE). 
Suppose we have $B$ point estimates of the AUC $\hat a^{(i)}$ for $i=1,...,B$. 
The RB in percentage and the MSE are respectively  defined to be:
\begin{equation*}
RB(\%)=\frac{1}{B}\sum_{i=1}^B \frac{\hat a^{(i)}-a_0}{a_0}\times100~~{\rm and}~~MSE=\frac{1}{B}\sum_{i=1}^B (\hat a^{(i)}-a_0)^2,
\end{equation*}
where $a_0$ is the true value of the AUC. 
The results are summarized in  Table \ref{table:auc}. 

{\blue Comparing the RB values in this table, we observe that our method is small in RB values for all examples, though in most cases, the RB values from the ECDF  method are the smallest; the kernel method always results in negative and largest RBs in absolute values. Comparing the MSE values, our BP method has comparable or better performance than competitive methods.}

\begin{table}[!htbp]
\caption{RB ($\%$) and MSE ($\times 1000$) of five methods for estimating the AUC}
\centering
\begin{tabular*}{\textwidth}{c@{\extracolsep{\fill}} c c c c c c c} 
\toprule
 &$(n_0,n_1)$ & \multicolumn{2}{c}{$(50,50)$} &  \multicolumn{2}{c}{$(100,100)$} &  \multicolumn{2}{c}{$(150,50)$} \\ 
 \cline{3-8}
Distribution ($J$)&Method &RB & MSE &RB & MSE&RB & MSE \\
\midrule
Normal (0.3) %&BP&-0.31&2.42&-0.31&1.23&-0.12&1.66\\
%&auc1&0.90&2.39&0.30&1.21&0.51&1.66\\
&BP&0.29&2.39&0.00&1.21&0.20&1.65\\
&Box-Cox & 0.44 & 2.44 & 0.06 &  1.23 & 0.20 & 1.68\\
 &ZL & 0.44 &2.47 &0.09 &1.24 &0.24& 1.69\\
%&LZL&0.09&2.51&-0.10&1.27&-0.01&1.74\\
& ECDF & 0.09 & 2.51 & -0.10 & 1.27 & -0.01 & 1.74\\
%& MNLE & 5.09 & 3.39 & 3.39 & 1.69 & 4.17 & 2.35\\
& Kernel & -1.67 & 2.38 & -1.54 & 1.28 & -1.53 & 1.71\\
%& MSLE &-1.35&2.23&-1.41&1.24&-1.33&1.63\\
\midrule
Normal (0.5) %&BP&-0.24&1.54&-0.21&0.76&-0.06&1.02\\
%&auc1&0.56&1.49&0.20&0.75&0.37&1.01\\
&BP&0.16&1.50&-0.01&0.75&0.16&1.01\\
&Box-Cox & 0.35 & 1.50 & 0.08 &  0.75 & 0.17 & 1.02\\
 &ZL & 0.36 &1.53& 0.10& 0.76& 0.19& 1.03\\
%&LZL&0.03&1.56&-0.07&0.77&-0.01&1.06\\
& ECDF & 0.03 & 1.56 & -0.07 & 0.77 & -0.01 & 1.06\\
%& MNLE & 2.90 & 1.88 & 1.95 & 0.94 & 2.37 & 1.29\\
& Kernel & -1.95 & 1.77 & -1.68 & 0.95 & -1.71 & 1.24\\
%& MSLE &-1.88&1.75&-1.66&0.95&-1.67&1.23\\
\midrule
Normal (0.7) %&BP&-0.18&0.60&-0.12&0.29&-0.05&0.38\\
%&auc1&0.27&0.57&0.10&0.28&0.20&0.37\\
&BP&0.05&0.58&-0.01&0.28&0.08&0.37\\
&Box-Cox & 0.10 & 0.54 & 0.01 &  0.27 & 0.05 & 0.36\\
&ZL & 0.14& 0.58 &0.04 &0.28 &0.10 &0.40\\
%&LZL&-0.02&0.61&-0.04&0.29&-0.01&0.39\\
& ECDF & -0.02 & 0.61 & -0.04 & 0.29 & -0.01 & 0.39 \\
%& MNLE & 1.40 & 0.63 & 1.00 & 0.33 & 1.18 & 0.43\\
& Kernel & -1.56 & 0.88 & -1.28 & 0.46 & -1.33 & 0.59 \\
%& MSLE & -1.54 & 0.87 & -1.27 & 0.46 & 1.31 & 0.59\\
\midrule
Gamma (0.3) %&BP&0.92&2.50&-0.01&1.26&-0.20&1.71\\
%&auc1&1.29&2.52&0.60&1.26&0.43&1.70\\
&BP&0.69&2.49&0.29&1.26&0.11&1.70\\
&Box-Cox&0.57&2.58&0.26&1.30&0.23&1.76\\
&ZL&0.55&2.60&0.24&1.30&0.20&1.75\\
%&LZL&0.05&2.66&-0.08&1.35&-0.10&1.78\\
&ECDF&0.04&2.66&-0.08&1.34&-0.10&1.78\\
%&MNLE&5.04&3.47&3.40&1.77&4.04&2.34\\
&Kernel&-2.36&2.63&-2.08&1.44&-2.28&1.86\\
%&MSLE&-1.88&2.43&-1.85&1.38&-1.96&1.75\\
\midrule
Gamma (0.5) %&BP&0.06&1.59&-0.14&0.80&-0.05&0.95\\
%&auc1&0.73&1.56&0.26&0.79&0.38&0.94\\
&BP&0.33&1.56&0.06&0.80&0.17&0.94\\
&Box-Cox&0.40&1.55&0.13&0.80&0.39&0.95\\
&ZL&0.39&1.58&0.10&0.81&0.36&0.97\\
%&LZL&-0.01&1.65&-0.14&0.84&0.13&0.99\\
&ECDF&-0.01&1.65&-0.14&0.84&0.13&0.99\\
%&MNLE&2.85&1.92&1.88&0.99&2.43&1.25\\
&Kernel&-2.65&2.12&-2.31&1.20&-2.33&1.37\\
%&MSLE&-2.48&2.04&-2.22&1.17&-2.23&1.33\\
\midrule
Gamma (0.7) %&BP&-0.12&0.61&-0.12&0.31&-0.12&0.35\\
%&auc1&0.32&0.58&0.10&0.30&0.13&0.34\\
&BP&0.10&0.59&-0.01&0.30&0.00&0.35\\
&Box-Cox&0.08&0.55&-0.01&0.29&0.08&0.32\\
&ZL&0.13&0.59&-0.01&0.30&0.14&0.39\\
%&LZL&-0.05&0.62&-0.10&0.31&0.00&0.35\\
&ECDF&-0.05&0.62&-0.10&0.31&0.00&0.35\\
%&MNLE&1.35&0.63&0.94&0.34&1.14&0.40\\
&Kernel&-2.19&1.14&-1.82&0.65&-2.03&0.80\\
%&MSLE&-2.10&1.11&-1.77&0.63&-1.98&0.78\\
\bottomrule
\end{tabular*}
\label{table:auc}
\end{table}

\subsection{Comparison of the Youden index and optimal cutoff point estimation}
In this section, we compare our method with the competitive methods in the estimation of the Youden index and the optimal cutoff point.
The criteria for comparison are the RB and MSE, which are similarly  defined as those in Section \ref{simu.auc}. 
The simulation results for estimating the Youden index and the optimal cutoff point are summarized in Tables \ref{table:youden} and \ref{table:cutoff}, respectively. 

{\blue From Table \ref{table:youden}, we observe that in the estimation of the Youden index, out of 18 simulation settings, our BP method leads to the smallest RB values in 16 settings, and obtains the second smallest RB values in 2 settings, where the kernel method produces the smallest RB values. Comparing MSE values, the performance of our BP method and the Box-Cox method is comparable, and is better than the other methods in most of the settings.}

{\blue From Table \ref{table:cutoff}, for estimating the cutoff point, our BP method and the Box-Cox method result in significantly smaller MSE values than other methods. The performance of the BP and the Box-Cox method is mixed. When $f_0$ and $f_1$ are simulated from the Gamma distribution and the Youden index is 0.7, Box-Cox method produces smaller MSE values, whereas our BP method leads to smaller MSE values in all other settings. }

\begin{table}[!http]
\caption{RB ($\%$) and MSE ($\times 1000$) of five methods for estimating the Youden index}
\centering
\begin{tabular*}{\textwidth}{c@{\extracolsep{\fill}} c c c c c c c} 
\toprule
 &$(n_0,n_1)$ & \multicolumn{2}{c}{$(50,50)$} &  \multicolumn{2}{c}{$(100,100)$} &  \multicolumn{2}{c}{$(150,50)$} \\ 
 \cline{3-8}
Distribution ($J$)&Method &RB & MSE &RB & MSE&RB & MSE \\
\hline
Normal (0.3) &BP&1.90&5.69&0.44&2.87&1.25&3.93\\
&Box-Cox & 4.25 & 5.86 & 1.57 &  2.89 & 2.50 & 3.97\\
&ZL & 4.66 &5.94& 1.80 &2.92& 2.84& 4.01\\
%&LZL&24.04&11.63&15.55&5.57&19.03&7.75\\
& ECDF & 24.05 & 11.64 & 11.57 & 5.57 & 19.05 & 7.75\\
%& MNLE & 24.05 & 11.64 & 11.57 & 5.57 & 19.05 & 7.75\\
& Kernel & -0.88 & 5.61 & -2.53 & 3.08 & -1.83 & 4.05\\
%& MSLE &-0.88&5.61&-2.53&3.08&-1.83&4.05\\
\midrule
Normal (0.5) &BP&1.32&4.91&0.40&2.43&0.99&3.29\\
&Box-Cox & 2.64 & 5.04 & 1.05 &  2.42 & 1.52 & 3.33\\
&ZL & 2.89 &5.24 &1.15& 2.47 &1.70 &3.40\\
%&LZL&11.66&9.15&7.39&4.33&9.27&6.12\\
& ECDF & 11.67 & 9.15 & 7.40 & 4.33 & 9.28 & 6.13\\
%& MNLE & 11.67 & 9.15 & 7.40 & 4.33 & 9.28 & 6.13\\
& Kernel & -3.53 & 5.45 & -3.73 & 3.01 & -3.54 & 3.90\\
%& MSLE & -3.53 & 5.45 & -3.73 & 3.01 & -3.54 & 3.90\\
\midrule
Normal (0.7) &BP&1.01&3.51&0.37&1.68&0.74&2.25\\
& Box-Cox & 1.64 & 3.38 & 0.70 &  1.61 & 0.95 & 2.20\\
&ZL & 2.09 &3.84 &0.89 &1.77 &1.40&2.87\\
%&LZL&6.28&5.70&4.06&2.79&5.00&3.79\\
& ECDF & 6.29 & 5.71 & 4.06 & 2.79 & 5.01 & 3.80\\
%& MNLE & 6.29 & 5.71 & 4.06 & 2.79  & 5.01 & 3.80\\
& Kernel & -3.74 & 4.43 & -3.43 & 2.43 & -3.43 & 3.10\\
%& MSLE & -3.74 & 4.43 & -3.43 & 2.43 & -3.43 & 3.10\\
\midrule
Gamma (0.3)&BP &3.85&6.03&1.74&2.98&1.32&4.03\\
&Box-Cox &5.27&6.25&2.90&3.08&3.18&4.27\\
&ZL&5.60&6.32&2.92&3.08&3.19&4.23\\
%&LZL&23.76&11.65&15.75&5.82&19.07&7.84\\
&ECDF &23.77&11.66&15.77&5.83&19.08&7.85\\
%&MNLE&23.77&11.66&15.77&5.83&19.08&7.85\\
&Kernel&-0.67&6.10&-1.91&3.35&-0.95&4.17\\
%&MSLE&-0.67&6.10&-1.91&3.35&-0.95&4.17\\
\midrule
Gamma (0.5) & BP&1.92&5.27&0.57&2.56&1.08&3.09\\
&Box-Cox&3.26&5.33&1.66&2.58&2.62&3.33\\
&ZL&3.30&5.47&1.54&2.60&2.43&3.35\\
%&LZL&11.19&9.01&7.24&4.45&9.37&5.94\\
&ECDF &11.20&9.02&7.25&4.45&9.38&5.94\\
%&MNLE&11.20&9.02&7.25&4.45&9.38&5.94\\
&Kerel&-3.42&5.91&-3.64&3.21&-2.96&3.55\\
%&MSLE&-3.42&5.91&-3.64&3.21&-2.96&3.55\\
\midrule
Gamma (0.7) &BP&1.34&3.66&0.51&1.83&0.71&2.09\\
&Box-Cox&1.88&3.46&0.95&1.75&1.37&2.08\\
&ZL&2.28&4.03&1.01&1.89&1.81&3.26\\
%&LZL&6.16&5.87&3.85&2.86&4.95&3.62\\
&ECDF &6.17&5.87&3.85&2.86&4.95&3.62\\
%&MNLE&6.17&5.87&3.85&2.86&4.95&3.62\\
&Kernel&-3.71&4.56&-3.50&2.62&-3.62&2.88\\
%&MSLE&-3.71&4.56&-3.50&2.62&-3.62&2.88\\
\bottomrule
\end{tabular*}
\label{table:youden}
\end{table}

\begin{table}[!http]
\caption{RB ($\%$) and MSE ($\times 1000$) of five methods for estimating the optimal cutoff point}
\centering
\begin{tabular*}{\textwidth}{c@{\extracolsep{\fill}} c c c c c c c} 
\toprule
 &$(n_0,n_1)$ & \multicolumn{2}{c}{$(50,50)$} &  \multicolumn{2}{c}{$(100,100)$} &  \multicolumn{2}{c}{$(150,50)$} \\ 
 \cline{3-8}
Distribution ($J$)&Method &RB & MSE &RB & MSE&RB & MSE \\
\midrule
Normal (0.3) &BP&-0.19&11.34&-0.18&5.72&-0.17&7.60\\
&Box-Cox & -0.11 & 44.14 & -0.02 &  22.61 & -0.08 & 29.02\\
&ZL&-0.22&58.06&-0.04&27.42&-0.05&35.82\\
%&LZL&-0.17&146.41&-0.10&100.75&0.16&118.25\\
& ECDF & -0.14 & 137.99 & -0.13 & 95.68 & 0.15 & 114.03\\
%& MNLE & -0.13 & 137.31 & -0.13 & 95.25 & 0.14 & 113.75\\
& Kernel & 0.04 & 110.52 & 0.06 & 64.31 & 0.28 & 79.47\\
%& MSLE &0.04&110.52&-0.06&64.32&0.28&79.45\\
\midrule
Normal (0.5) &BP&-0.15&11.40&-0.13&5.70&-0.11&7.51\\
&Box-Cox & -0.09 & 22.44 & -0.02 &  11.15 & -0.03 & 13.51\\
&ZL&-0.22&33.61&-0.06&15.88&0.00&19.40\\
%&LZL&-0.25&87.21&-0.12&59.57&0.13&70.82\\
& ECDF & -0.22 & 82.58 & -0.10 & 56.55 & -0.14 & 68.46 \\
%& MNLE & -0.22 & 82.46 & -0.10 & 56.07 & -0.14 & 68.40\\
& Kernel & -0.03 & 42.51 & -0.04 & 23.82  & 0.09 & 30.02\\
%& MSLE  &-0.04&42.52&-0.04&23.82&0.08&30.02\\
\midrule
Normal (0.7) &BP&-0.11&12.96&-0.10&6.37&-0.05&8.44\\
&Box-Cox & -0.06 & 15.56 & -0.02 &  7.48 & 0.01 & 9.23\\
&ZL &-0.23&31.01&-0.08&14.66&-0.01&18.19\\
%&LZL&-0.29&64.95&-0.08&40.22&0.15&52.01\\
& ECDF & -0.27 & 62.74 & -0.10 & 38.93 & 0.13 & 51.15\\
%& MNLE & -0.26 & 62.73 & -0.09 & 38.80 & 0.13 & 51.15\\
& Kernel & -0.04 & 27.67 & 0.01 & 14.65 & 0.03 & 19.03 \\
%& MSLE &-0.05&27.67&0.00&14.65&0.02&19.03\\
\midrule
Gamma (0.3) &BP&4.48&58.33&2.96&31.53&3.67&42.28\\
&Box-Cox&0.21&88.72&-0.38&43.95&-0.42&52.51\\
&ZL&0.44&120.92&-0.12&56.77&0.67&71.78\\
%&LZL&2.81&341.48&1.94&218.63&4.09&284.06\\
&ECDF&2.86&319.78&1.78&206.42&4.09&274.15\\
%&MNLE&2.67&313.56&1.65&204.17&4.00&272.66\\
&Kernel&12.12&307.63&8.73&150.92&10.95&189.00\\
%&MSLE&12.10&307.44&8.71&150.78&10.93&188.82\\
\midrule
Gamma (0.5) &BP&2.69&45.48&1.98&24.47&2.48&32.88\\
&Box-Cox&-0.46&52.60&-0.42&27.26&-0.27&30.60\\
&ZL&-0.39&81.87&-0.07&39.84&0.87&47.04\\
%&LZL&-0.14&227.38&0.49&149.22&3.86&211.86\\
&ECDF&0.01&210.50&0.62&144.56&2.08&181.49\\
%&MNLE&-0.03&209.13&0.60&144.27&2.03&180.81\\
&Kernel&7.17&141.60&5.72&80.26&6.97&94.56\\
%&MSLE&7.16&141.44&5.70&80.14&6.95&94.40\\
\midrule
Gamma (0.7) & BP&1.53&57.16&1.15&31.92&1.53&41.74\\
&Box-Cox&-0.62&51.57&-0.57&26.03&-0.42&32.98\\
&ZL&-0.59&106.81&-0.24&53.14&0.39&66.29\\
%&LZL&-0.70&228.96&0.08&156.61&1.41&198.65\\
&ECDF&-0.59&223.80&-0.12&150.58&1.27&193.70\\
%&MNLE&-0.62&222.60&-0.12&149.71&1.22&192.81\\
&Kernel&4.38&140.67&3.56&75.42&4.03&82.75\\
%&MSLE&4.36&140.51&3.54&75.29&4.01&82.60\\
\bottomrule
\end{tabular*}
\label{table:cutoff}
\end{table}

\section{Real data application}
\label{realdata}

In this section, we apply our method to a dataset on Duchenne Muscular Dystrophy (DMD). 
DMD is a type of muscular dystrophy that is genetically transmitted from a mother to her children.  
Muscle loss occurs at an early age for offspring with the disease.
 Female offspring with the disease do not suffer from significant symptoms compared to male offspring who die at a young age.
  Female carriers do not show sign of disease and therefore detection of potential female carrier is of main interest.

\cite{Percy1982} stated that DMD carriers are more likely to have higher measurement of specific biomarkers. 
Four biomarkers including creatine kinase (CK), hemopexin
(H), lactate dehydroginase (LD), and pyruvate kinase (PK) are measured from the blood serum samples of a healthy group ($n_0=127$) and a group of carriers ($n_1=67$).
The complete dataset was collected by  \cite{Andrews2012}.
\cite{Yuan2021} pointed out that the biomarker CK has the best performance among these four biomarkers since it corresponds to largest estimate of the Youden index. 
Therefore, we apply our method and other existing methods on the biomarker CK to compare their performance.

{\blue Following Remark \ref{remark-1}, we construct the $BIC(N)$ versus $N$ plot in Panel (a) of Figure \ref{BICCDF}; it suggests to use $N = 1$ in our BP method. Based on Remark \ref{remark-4}, to check whether the likelihood ratio ordering assumption is reasonable, we plot our BP estimates $\left(\hat F_0,\hat F_1\right)$ and the empirical cdf estimates $\left(\tilde F_0, \tilde F_1\right)$ in Panel (b) of Figure \ref{BICCDF}; this plot suggests that the likelihood ratio ordering assumption might be reasonable. 
We further perform the goodness-of-fit test suggested in Remark \ref{remark-4}; the $p$-value based on 1000 bootstrap sample is 0.975. This reinforces the validity of the  likelihood ratio ordering assumption.} 

\begin{figure}[!ht]
 \begin{center}
 \includegraphics[scale=0.45]{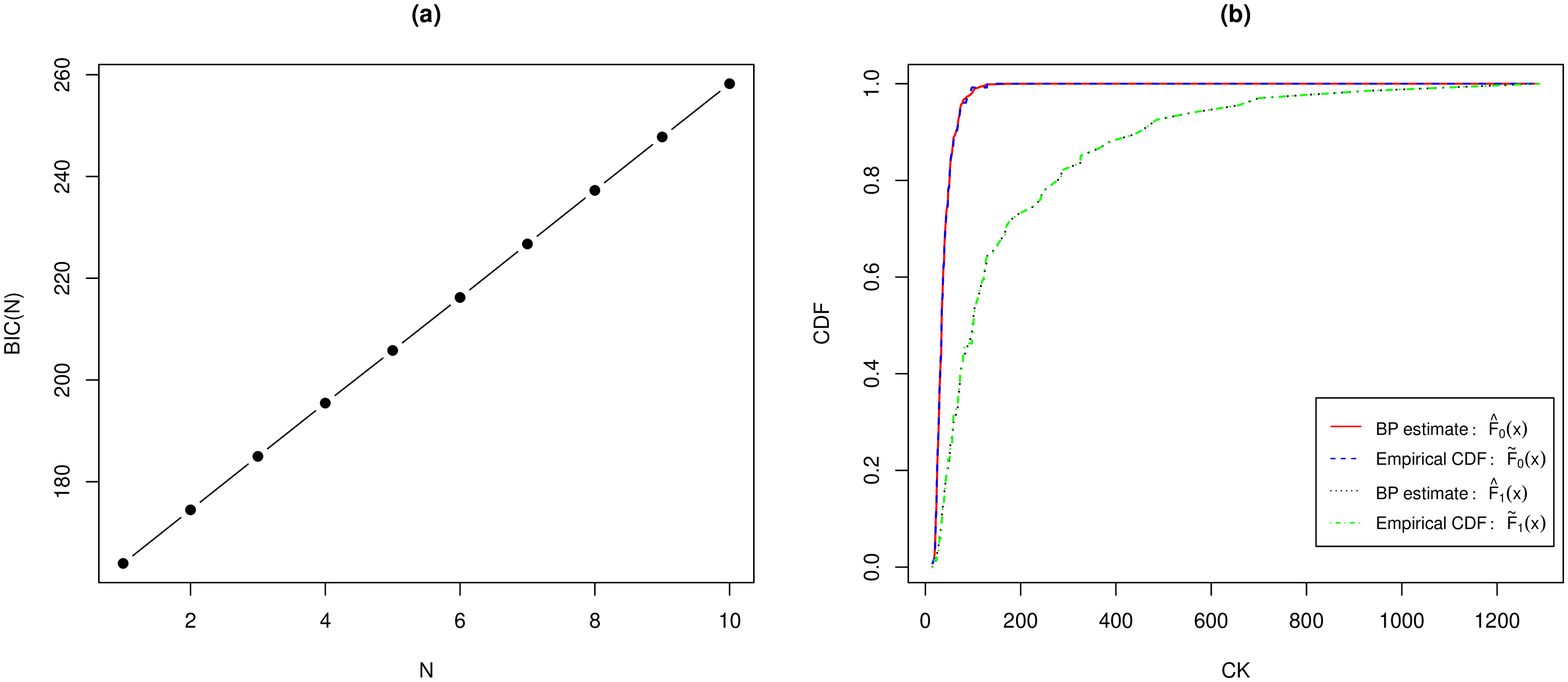}
\end{center}
\caption{
Panel (a) shows $BIC(N)$ versus $N$; Panel (b) compares the proposed estimators $\big(\hat F_0,\hat F_1\big)$
and the empirical cdfs $\big(\tilde F_0, \tilde F_1\big)$. 
}
\label{BICCDF}
\end{figure}

{\blue 
In Figure \ref{Figure: ROC curve}, we plot the ROC curve estimates from our BP and competitive methods. The performance of the MNLE and MSLE methods is similar to that of the ECDF and kernel method, respectively; we thus do not include them in this figure. From the figure, we observe that the ROC curve estimate from our BP method is similar to those of Box-Cox, ZL, ECDF, and LZL methods, but different from the kernel method. The difference lies in the fact that the kernel method leads to much smaller sensitivity estimate compared to other methods when 1-specificity is large, greater than 0.5 say. This is mainly because that the CK values for the diseased individuals are very skewed, which inflates the selected bandwidth when the kernel method is used to estimate $F_1$; this in turn makes $F_1$ overestimated and hence causes the sensitivity underestimated when 1-specificity is large or the value of CK is small. 
}

\begin{figure}[!ht]
 \begin{center}
 \includegraphics[scale=0.45]{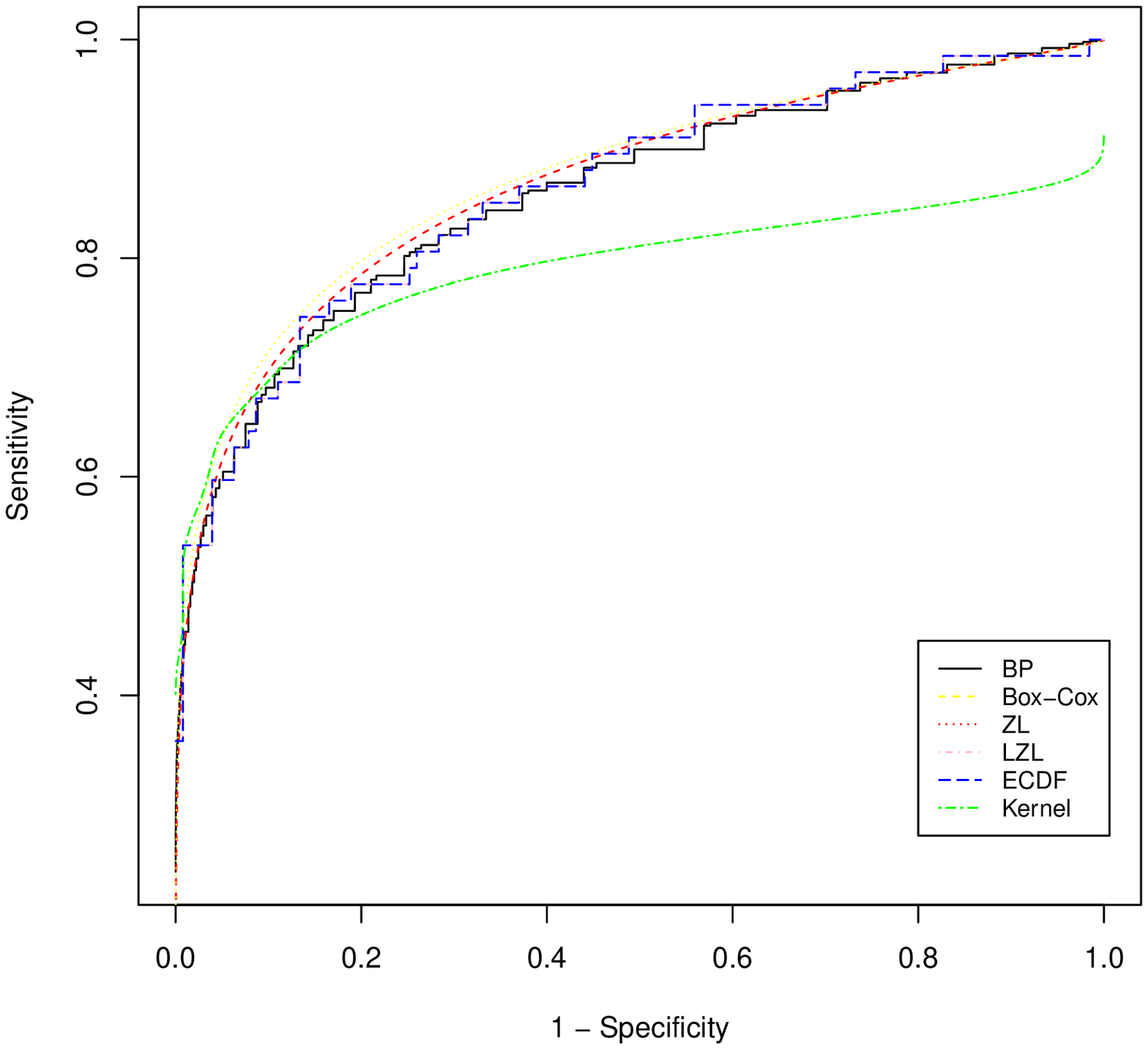}
\end{center}
\caption{
Estimated ROC curves based on different methods. 
}
\label{Figure: ROC curve}
\end{figure}

{\blue We evaluate the point estimates (PEs) and bootstrap percentile confidence intervals (BPCIs) of the AUC, Youden index, and optimal cutoff point based on different methods, and display them in Table \ref{table:real}. The AUC estimates from all methods are similar, except for the kernel method, which leads to a smaller AUC estimate. This is consistent with the results in Figure \ref{Figure: ROC curve}, and complies with our observation in Table \ref{table:auc}: the kernel method has the negative relative biases, and all other methods have small RBs and comparable MSEs. The Youden index estimates are similar for all methods: the maximum difference between any two estimates is less than 0.03. For the optimal cutoff point estimates, the kernel method leads to a different result from other methods; other methods give similar results. But the ZL, LZL and ECDF methods have resulted in wider BPCIs than our method; we conjecture that this may indicate that our method has 
given smaller standard error than other methods in the optimal cutoff point estimation. However, unlike with the simulation studies, in this real data example, we are unable to compare the estimates with the true values. 
}

\begin{table}[!http]
\caption{PEs and BPCIs of the AUC, Youden index, and optimal cutoff point by using CK as the biomarker}
\bigskip
\centering
\begin{tabular*}{\textwidth}{c@{\extracolsep{\fill}} c c c c c c} 
\toprule
Statistics&\multicolumn{2}{c}{AUC}& \multicolumn{2}{c}{Youden index} & \multicolumn{2}{c}{Optimal cutoff point}\\
Method & PE & BPCI & PE & BPCI & PE & BPCI \\
\midrule
BP  &0.865  & [0.804, 0.914]& 0.588 & [0.480, 0.688] & 58.998 & [51.976, 67.311]\\
Box-Cox&  0.874 & [0.816, 0.921]  &0.616 & [0.510, 0.710] & 58.017 & [51.869, 65.055] \\
ZL & 0.868 & [0.805, 0.917] & 0.600 & [0.498, 0.697]& 57.000 & [50.975, 69.000] \\
LZL & 0.863 &[0.800, 0.914] & 0.612 & [0.502, 0.727] & 56.000 & [43.000, 97.500] \\
ECDF &  0.863 & [0.800, 0.914] &0.612 & [0.502, 0.727]  & 56.000 & [43.000, 97.500] \\
Kernel &  0.790 & [0.748, 0.844] &0.591 &[0.502, 0.669]  &73.356 & [54.689, 79.413] \\
\bottomrule
\end{tabular*}
\label{table:real}
\end{table}

\section{Conclusion}
\label{conclude}

We have proposed a Bernstein polynomial method to estimate the cdfs of the biomarkers in the two-sample problem under the likelihood ratio ordering assumption, and subsequently considered the 
the estimation of the ROC curve and its summary statistics. With simulation studies, we have compared our method with existing methods; we observe that our method performs well in the estimation of both the ROC curve and the summary statistics. 
{\blue With the empirical process theory, we established the consistency of our estimators.}
We implemented the numerical algorithm for our method in an R package named {\tt BPLR}, which is ready to be applied in practice. 

{\blue We observe that within the framework of this article, there are many interesting topics that can be explored in the future. First, we have established the consistency of our proposed estimators, but not the convergence rates and the asymptotic distributions; if we are able to establish the asymptotic distributions of them in the future, we can construct the confidence bands/intervals for $F_0(\cdot)$, $F_1(\cdot)$, $ROC(\cdot)$, and other summary statistics. 
Second, we have proposed to employ the Youden index as the criterion, and maximise it to determine the optimal cutoff point; however, there may exist other criteria; for example, the closest to (0,1) criterion and equal sensitivity and specificity criterion \citep{sande2021statistical}. It could be interesting to explore how to extend our method to accommodate these criteria. Third, in Remark \ref{remark-3}, we have suggested to include the log transformed biomarker in our model and have applied this strategy in both the simulation and real data analysis. We observe that there may exist other reasonable transformations that can be incorporated in our method, for example, \cite{yang2021transformation} proposed a transformation method based on the likelihood ratio and discussed the scenarios under which that transformation is applicable. 
We leave this for future research. 
}

%\subsection*{Conflict of interest}

%The authors have no conflicts of interest to declare.

%\appendix
{\blue
\section*{Appendix: Regularity Conditions}
We need the following conditions in our technical developments for Theorems \ref{thm1} and \ref{theorem-2}. They are not necessarily the weakest possible. 
\begin{enumerate}
    \item[A1] The total sample size $n=n_0+n_1\to\infty$ and 
    $n_1/n\to\lambda$ for $\lambda\in(0,1)$ being a constant. 
    \item[A2] The order $N\to\infty$ as $n\to\infty$ and $\lim_{n\to\infty} N/n=0$. 
    
    \item[A3] The likelihood ratio ordering assumption is satisfied, or equivalently, $\theta_0(x)$ in \eqref{def.theta0} is a nondecreasing function of $x$. 
    Furthermore, $\theta_0(x)$ is a continuous function of $x$ and there exists a $\delta>0$ such that $\delta\leq\theta_0(x)\leq 1-\delta$ for all $x\in[0,1]$. 
    \item [A4] There exists an $\epsilon_0>0$, such that $\theta_0(x)$ is strictly increasing for $x\in [C-\epsilon_0,C+\epsilon_0]$, where $C$ is defined by \eqref{def.cutoff}. 
\end{enumerate}
}

\newpage
\setcounter{section}{0}
\setcounter{equation}{0}
\setcounter{table}{0}

\def\theequation{S\arabic{equation}}
\def\thesection{S\arabic{section}}
\def\thetable{S\arabic{table}}

{\centering {\Large {Supplementary Material for \\ ``Statistical inference for the two-sample problem under likelihood ratio ordering, with application to the ROC curve estimation"}}}
\bigskip

\noindent

% \vspace{-1.5cm}
\centerline{\bf Abstract}
This is a supplementary document to the paper ``Statistical inference for the two-sample problem under likelihood ratio ordering, with application to the ROC curve estimation". Section \ref{supp-section-1} gives some additional simulation results. Section \ref{supp-section-2} proves Proposition 1 in the main article. Sections \ref{supp-section-3} and \ref{supp-section-4} present technical details for Theorems 1 and 2 in the main article. 

\setcounter{section}{0}

\section{Additional simulation results} \label{supp-section-1}

\subsection{Additional simulation results when $f_0$ and $f_1$ are from the normal and Gamma distributions}
In this section, we first present the frequency of the selected $N$ values based on the BIC criterion in Remark 1 of the main article; the results are given in Table \ref{table:N1}. 
Based on Remark 3 of the main article, we include both the original and log-transformed biomarkers in the model. Hence, for both normal and Gamma distributions, the true value of $N$ is 1. From Table \ref{table:N1}, we observe that the proposed BIC criterion can choose $N=1$ for the majority of repetitions in all the simulation settings. 

Tables \ref{table:roc1}--\ref{table:cutoff1} summarize the results of the maximum nonparametric likelihood method (MNLE), the maximum smoothed likelihood method (MSLE), and the method in \cite{lin2012direct} (LZL)
in the estimation of the receiver operating characteristic (ROC) curve, the area under the curve (AUC), 
the Youden index, and the optimal cutoff point. From these tables, we observe that
the performance of MNLE is similar to that of the empirical cumulative distribution function (ECDF) based method; the performance of the MSLE is close to that of the kernel method. The results of LZL are very similar to those of the ECDF method. 

\begin{table}[!htbp]
\caption{Frequency of the $N$ values when distributions are normal and Gamma}
\centering
\small
\begin{tabular*}{\textwidth}{c@{\extracolsep{\fill}} c c c c c c c c c c} 
\toprule
 & \multicolumn{3}{c}{ $(n_0,n_1)=(50,50)$} &  \multicolumn{3}{c}{ $(n_0,n_1)=(100,100)$} &  \multicolumn{3}{c}{ $(n_0,n_1)=(150,50)$} \\ 
 \cline{2-10}

Distribution ($J$)~~  & $N=1$ & $N=2$ & $N=3$ & $N=1$ & $N=2$ & $N=3$ & $N=1$ & $N=2$ & $N=3$ \\
\midrule
Normal ($0.3$)&1994&6&0&1999&1&0&1995&5&0\\
Normal ($0.5$)&1999&1&0&1999&1&0&1999&1&0\\
Normal ($0.7$)&2000&0&0&2000&0&0&2000&0&0\\
\midrule
Gamma ($0.3$)&1982&18&0&1986&14&0&1995&5&0\\
Gamma ($0.5$)&1988&12&0&1991&9&0&1992&8&0\\
Gamma ($0.7$)&1996&4&0&2000&0&0&2000&0&0\\
\bottomrule
\end{tabular*}
\label{table:N1}
\end{table}

\begin{table}[!htbp]
\caption{Averages of $L_1$- and $L_2$-distances of the ROC curve estimates}
\centering
\small
\begin{tabular*}{\textwidth}{c@{\extracolsep{\fill}} c c c c c c c} 
\toprule
 &$(n_0,n_1)$ & \multicolumn{2}{c}{$(50,50)$} &  \multicolumn{2}{c}{$(100,100)$} &  \multicolumn{2}{c}{$(150,50)$} \\ 
 \cline{3-8}

Distribution ($J$)&Method & $L_1$ & $L_2$ & $L_1$& $L_2$ & $L_1$ & $L_2$ \\
\midrule
Normal (0.3)& MNLE & 0.054 & 0.066 & 0.038 & 0.047 & 0.045 & 0.054 \\
& MSLE & 0.045 & 0.054 & 0.035 & 0.041 & 0.039 & 0.046\\
 &LZL&0.056&0.071&0.040&0.051&0.046&0.058\\
Normal (0.5)& MNLE & 0.042 & 0.059 & 0.030 & 0.043 & 0.035 & 0.047\\
& MSLE & 0.039 & 0.052 & 0.029 & 0.039 & 0.033 & 0.043\\
&LZL&0.045&0.064&0.032&0.046&0.037&0.051\\
Normal (0.7)& MNLE & 0.026 & 0.047 & 0.019 & 0.035 & 0.021 & 0.036 \\
& MSLE & 0.027 & 0.044 & 0.020 & 0.033 & 0.022 & 0.035\\
&LZL&0.029&0.053&0.020&0.038&0.023&0.040\\
\midrule
Gamma (0.3)& MNLE&0.054&0.067&0.039&0.048&0.044&0.054\\
&MSLE&0.049&0.059&0.037&0.044&0.042&0.049\\
&LZL&0.057&0.072&0.040&0.052&0.046&0.058\\
Gamma (0.5) &MNLE&0.042&0.061&0.031&0.044&0.034&0.047\\
&MSLE&0.045&0.058&0.034&0.044&0.037&0.046\\
&LZL&0.045&0.067&0.032&0.047&0.036&0.050\\
Gamma (0.7) &MNLE&0.025&0.049&0.019&0.036&0.020&0.036\\
&MSLE&0.033&0.049&0.025&0.037&0.027&0.038\\
&LZL&0.029&0.055&0.021&0.040&0.022&0.040\\
\bottomrule
\end{tabular*}
\label{table:roc1}
\end{table}

\begin{table}[!htbp]
\caption{RB ($\%$) and MSE ($\times 1000$) of the AUC estimates}
\centering
\small
\begin{tabular*}{\textwidth}{c@{\extracolsep{\fill}} c c c c c c c} 
\toprule
 &$(n_0,n_1)$ & \multicolumn{2}{c}{$(50,50)$} &  \multicolumn{2}{c}{$(100,100)$} &  \multicolumn{2}{c}{$(150,50)$} \\ 
 \cline{3-8}
Distribution ($J$)&Method &RB & MSE &RB & MSE&RB & MSE \\
\midrule
Normal (0.3)& MNLE & 5.09 & 3.39 & 3.39 & 1.69 & 4.17 & 2.35\\
& MSLE &-1.35&2.23&-1.41&1.24&-1.33&1.63\\
&LZL&0.09&2.51&-0.10&1.27&-0.01&1.74\\
Normal (0.5)& MNLE & 2.90 & 1.88 & 1.95 & 0.94 & 2.37 & 1.29\\
& MSLE &-1.88&1.75&-1.66&0.95&-1.67&1.23\\
&LZL&0.03&1.56&-0.07&0.77&-0.01&1.06\\
Normal (0.7)& MNLE & 1.40 & 0.63 & 1.00 & 0.33 & 1.18 & 0.43\\
& MSLE & -1.54 & 0.87 & -1.27 & 0.46 & 1.31 & 0.59\\
&LZL&-0.02&0.61&-0.04&0.29&-0.01&0.39\\
\midrule
Gamma (0.3) &MNLE&5.04&3.47&3.40&1.77&4.04&2.34\\
&MSLE&-1.88&2.43&-1.85&1.38&-1.96&1.75\\
&LZL&0.05&2.66&-0.08&1.35&-0.10&1.78\\
Gamma (0.5) &MNLE&2.85&1.92&1.88&0.99&2.43&1.25\\
&MSLE&-2.48&2.04&-2.22&1.17&-2.23&1.33\\
&LZL&-0.01&1.65&-0.14&0.84&0.13&0.99\\
Gamma (0.7) &MNLE&1.35&0.63&0.94&0.34&1.14&0.40\\
&MSLE&-2.10&1.11&-1.77&0.63&-1.98&0.78\\
&LZL&-0.05&0.62&-0.10&0.31&0.00&0.35\\
\bottomrule
\end{tabular*}
\label{table:auc1}
\end{table}

\begin{table}[!htbp]
\caption{RB ($\%$) and MSE ($\times 1000$) of the Youden index estimates}
\centering
\small
\begin{tabular*}{\textwidth}{c@{\extracolsep{\fill}} c c c c c c c} 
\toprule
 &$(n_0,n_1)$ & \multicolumn{2}{c}{$(50,50)$} &  \multicolumn{2}{c}{$(100,100)$} &  \multicolumn{2}{c}{$(150,50)$} \\ 
 \cline{3-8}

Distribution ($J$)&Method & $L_1$ & $L_2$ & $L_1$& $L_2$ & $L_1$ & $L_2$ \\
\midrule
Normal (0.3) & MNLE & 24.05 & 11.64 & 11.57 & 5.57 & 19.05 & 7.75\\
& MSLE &-0.88&5.61&-2.53&3.08&-1.83&4.05\\
&LZL&24.04&11.63&15.55&5.57&19.03&7.75\\
Normal (0.5)& MNLE & 11.67 & 9.15 & 7.40 & 4.33 & 9.28 & 6.13\\
& MSLE & -3.53 & 5.45 & -3.73 & 3.01 & -3.54 & 3.90\\
&LZL&11.66&9.15&7.39&4.33&9.27&6.12\\
Normal (0.7)& MNLE & 6.29 & 5.71 & 4.06 & 2.79  & 5.01 & 3.80\\
& MSLE & -3.74 & 4.43 & -3.43 & 2.43 & -3.43 & 3.10\\
&LZL&6.28&5.70&4.06&2.79&5.00&3.79\\
\midrule
Gamma (0.3)&MNLE&23.77&11.66&15.77&5.83&19.08&7.85\\
&MSLE&-0.67&6.10&-1.91&3.35&-0.95&4.17\\
&LZL&23.76&11.65&15.75&5.82&19.07&7.84\\
Gamma (0.5)&MNLE&11.20&9.02&7.25&4.45&9.38&5.94\\
&MSLE&-3.42&5.91&-3.64&3.21&-2.96&3.55\\
&LZL&11.19&9.01&7.24&4.45&9.37&5.94\\
Gamma (0.7) &MNLE&6.17&5.87&3.85&2.86&4.95&3.62\\
&MSLE&-3.71&4.56&-3.50&2.62&-3.62&2.88\\
&LZL&6.16&5.87&3.85&2.86&4.95&3.62\\
\bottomrule
\end{tabular*}
\label{table:Youden1}
\end{table}

\newpage

\begin{table}[!http]
\caption{RB ($\%$) and MSE ($\times 1000$) of the optimal cutoff point estimates}
\centering
\small
\begin{tabular*}{\textwidth}{c@{\extracolsep{\fill}} c c c c c c c} 
\toprule
 &$(n_0,n_1)$ & \multicolumn{2}{c}{$(50,50)$} &  \multicolumn{2}{c}{$(100,100)$} &  \multicolumn{2}{c}{$(150,50)$} \\ 
 \cline{3-8}
Distribution ($J$)&Method &RB & MSE &RB & MSE&RB & MSE \\
\midrule
Normal (0.3)& MNLE & -0.13 & 137.31 & -0.13 & 95.25 & 0.14 & 113.75\\
& MSLE &0.04&110.52&-0.06&64.32&0.28&79.45\\
&LZL&-0.17&146.41&-0.10&100.75&0.16&118.25\\
Normal (0.5)& MNLE & -0.22 & 82.46 & -0.10 & 56.07 & -0.14 & 68.40\\
& MSLE  &-0.04&42.52&-0.04&23.82&0.08&30.02\\
&LZL&-0.25&87.21&-0.12&59.57&0.13&70.82\\
Normal (0.7)& MNLE & -0.26 & 62.73 & -0.09 & 38.80 & 0.13 & 51.15\\
& MSLE &-0.05&27.67&0.00&14.65&0.02&19.03\\
&LZL&-0.29&64.95&-0.08&40.22&0.15&52.01\\
\midrule
Gamma (0.3) &MNLE&2.67&313.56&1.65&204.17&4.00&272.66\\
&MSLE&12.10&307.44&8.71&150.78&10.93&188.82\\
&LZL&2.81&341.48&1.94&218.63&4.09&284.06\\
Gamma (0.5) &MNLE&-0.03&209.13&0.60&144.27&2.03&180.81\\
&MSLE&7.16&141.44&5.70&80.14&6.95&94.40\\
&LZL&-0.14&227.38&0.49&149.22&3.86&211.86\\
Gamma (0.7) &MNLE&-0.62&222.60&-0.12&149.71&1.22&192.81\\
&MSLE&4.36&140.51&3.54&75.29&4.01&82.60\\
&LZL&-0.70&228.96&0.08&156.61&1.41&198.65\\
\bottomrule
\end{tabular*}
\label{table:cutoff1}
\end{table}

\subsection{Simulation results when $f_0$ and $f_1$ follow the Beta distribution}

In this section, we consider the simulation when $f_0\sim \text{Beta}(2, 2)$ and $f_1\sim \text{Beta}(a_1, b_1)$. For $(a_1, b_1)$, we consider three sets of values as shown in Table \ref{table:betapara} such that the corresponding $J$ is 0.3, 0.5, and 0.7 respectively. We note that under this simulation setup, the true value of $N$ is 1. 

Table \ref{table:N2} summarizes the frequency of the selected $N$ values based on the BIC criterion in Remark 1 of the main article; this table shows that the BIC criterion picks the correct order in the majority of the repetitions. 
Tables \ref{table:roc2}--\ref{table:cutoff2} summarize the results of all methods
for estimating the ROC  curve, the AUC, 
the Youden index, and the optimal cutoff point. 
The observations are similar to those when $f_0$ and $f_1$ are simulated as normal and Gamma distributions; we omit the details. 
It is noteworthy that the model assumption for the Box-cox method is violated under this setup. 
Our BP method performs better than the Box-Cox method for estimating the ROC curve, the Youden index, and the optimal cutoff point; the improvement is significant. 
For estimating the AUC, our BP method leads to smaller MSEs when $J=0.3$ and 0.5, but the Box-Cox method results in smaller MSEs when $J=0.7$.

\begin{table}[!http]
\caption{Beta simulation setting}
\centering
\small
\begin{tabular*}{\textwidth}{c@{\extracolsep{\fill}} c c   c c} 
\toprule
Distribution & $J$ & $AUC$ & $a_1$ & $b_1$  \\
\midrule
Beta &0.3 & 0.702 & 3.838 & 2\\
Beta &0.5 & 0.822 & 6.148 & 2\\
Beta &0.7 & 0.919 & 11.014 & 2\\
\bottomrule

\end{tabular*}
\label{table:betapara}
\end{table}

\begin{table}[!htbp]
\caption{Frequency of the chosen $N $ in Beta distributional setting}
\centering
\small
\begin{tabular*}{\textwidth}{c@{\extracolsep{\fill}} c c c c c c c c c c} 
\toprule
 $(n_0,n_1)$ & \multicolumn{3}{c}{$(50,50)$} &  \multicolumn{3}{c}{$(100,100)$} &  \multicolumn{3}{c}{$(150,50)$} \\ 
 \cline{2-10}

Distribution ($J$)~~ & $N=1$ & $N=2$ & $N=3$ & $N=1$ & $N=2$ & $N=3$ & $N=1$ & $N=2$ & $N=3$ \\
\midrule
Beta ($0.3$)&1989&11&0&1996&4&0&1998&2&0\\
Beta ($0.5$)&1999&1&0&2000&0&0&2000&0&0\\
Beta ($0.7$)&2000&0&0&2000&0&0&2000&0&0\\
\bottomrule
\end{tabular*}
\label{table:N2}
\end{table}

\newpage

\begin{table}[!htbp]
\caption{Averages of $L_1$-distances and $L_2$-distances of eight methods for estimating the ROC curve}
\centering
\small
\begin{tabular*}{\textwidth}{c@{\extracolsep{\fill}} c c c c c c c} 
\toprule
 &$(n_0,n_1)$ & \multicolumn{2}{c}{$(50,50)$} &  \multicolumn{2}{c}{$(100,100)$} &  \multicolumn{2}{c}{$(150,50)$} \\ 
 \cline{3-8}

Distribution ($J$)&Method & $L_1$ & $L_2$ & $L_1$& $L_2$ & $L_1$ & $L_2$ \\
\midrule
Beta (0.3) &BP&0.041&0.048&0.029&0.034&0.031&0.037\\
&Box-Cox & 0.045 & 0.055 & 0.032 &  0.039 & 0.036 & 0.043\\
&ZL & 0.046& 0.057& 0.032& 0.039& 0.035& 0.043\\
&LZL&0.056&0.073&0.040&0.052&0.044&0.057\\
& ECDF & 0.056 & 0.073 & 0.040 & 0.052 & 0.044 & 0.057\\
& MNLE & 0.054 & 0.069 & 0.039 & 0.050 & 0.044 & 0.055\\
& Kernel & 0.047 & 0.058 & 0.034 & 0.042 & 0.037 & 0.046 \\
& MSLE & 0.044 & 0.054 & 0.034 & 0.040 & 0.035 & 0.044\\
\midrule
Beta (0.5) &BP&0.033&0.046&0.024&0.033&0.024&0.033\\
&Box-Cox & 0.039 &0.055& 0.029& 0.041& 0.031& 0.044\\
&ZL & 0.038 &0.054 &0.027& 0.038& 0.027& 0.039\\
&LZL&0.046&0.070&0.032&0.050&0.034&0.051\\
& ECDF & 0.046 &0.070& 0.032& 0.050 &0.034 &0.051\\
& MNLE & 0.043& 0.065& 0.032 &0.047& 0.032& 0.047\\
& Kernel & 0.040& 0.057 &0.029 &0.042& 0.030 &0.042\\
& MSLE &  0.038 &0.054& 0.028& 0.040 &0.029 &0.041\\
\midrule
Beta (0.7) &BP&0.022&0.041&0.015&0.029&0.015&0.028\\
&Box-Cox & 0.028 &0.056 &0.023 &0.046 &0.025 &0.051\\
&ZL & 0.025 &0.050 &0.017 &0.034& 0.020& 0.038\\
&LZL&0.031&0.065&0.021&0.045&0.022&0.044\\
& ECDF & 0.031 &0.065 &0.021& 0.045& 0.022 &0.044\\
& MNLE & 0.028& 0.059 &0.020 &0.043& 0.020& 0.041\\
& Kernel & 0.030 &0.058& 0.022& 0.044 &0.022& 0.042\\
& MSLE &  0.028& 0.053 &0.021 &0.040 &0.021 &0.039\\
\bottomrule
\end{tabular*}
\label{table:roc2}
\end{table}

\newpage

\begin{table}[!htbp]
\caption{RB ($\%$) and MSE ($\times 1000$) of eight methods for estimating the AUC}
\centering
\small
\begin{tabular*}{\textwidth}{c@{\extracolsep{\fill}} c c c c c c c} 
\toprule
 &$(n_0,n_1)$ & \multicolumn{2}{c}{$(50,50)$} &  \multicolumn{2}{c}{$(100,100)$} &  \multicolumn{2}{c}{$(150,50)$} \\ 
 \cline{3-8}
Distribution ($J$)&Method &RB & MSE &RB & MSE&RB & MSE \\
\midrule
Beta (0.3)& BP &1.07&2.42&0.82&1.20&0.82&1.43\\
&Box-Cox & 1.12 & 2.72 & 0.99 &  1.35 & 1.17 & 1.61\\
&ZL & 0.62& 2.60 &0.50 &1.27 &0.51 &1.51\\
&LZL&0.14&2.65&0.13&1.31&0.18&1.53\\
& ECDF & 0.14 & 2.65 & 0.13 & 1.31 & 0.18 & 1.53 \\
& MNLE  & 5.33 & 3.57 & 3.74 & 1.82 & 4.35 & 2.22\\
& Kernel & -1.53 & 2.50 & -1.19 & 1.27 & -1.18 & 1.48\\
& MSLE & -1.08 & 2.30 & -0.98 & 1.20 & -0.93 & 1.38\\
\midrule
Beta (0.5) &BP&0.66&1.58&0.41&0.81&0.31&0.84\\
&Box-Cox & 1.48 &1.74& 1.30 &0.93& 1.16& 0.90\\
&ZL & 0.56& 1.68& 0.32& 0.85& 0.26& 0.87\\
&LZL&0.13&1.73&0.02&0.87&0.02&0.89\\
& ECDF & 0.13 &1.73& 0.02 &0.87 &0.02 &0.89\\
& MNLE & 3.16& 2.08 &2.16& 1.08& 2.37& 1.13\\
& Kernel & -1.79 & 1.86& -1.51&  0.99 &-1.46 & 1.00\\
& MSLE &  -1.57 & 1.76 &-1.38 & 0.95 &-1.36 & 0.97\\
\midrule
Beta (0.7) &BP&0.36&0.69&0.23&0.33&0.16&0.34\\
&Box-Cox & 0.58& 0.60& 0.50 &0.29& 0.09& 0.25\\
&ZL & 0.36 &0.73& 0.20 &0.36 &0.23& 0.83\\
&LZL&0.11&0.77&0.04&0.37&0.06&0.35\\
& ECDF & 0.11&0.77& 0.04 &0.37& 0.06 &0.35\\
& MNLE & 1.70& 0.83& 1.20& 0.43& 1.23 &0.42\\
& Kernel & -1.79&  1.10& -1.48&  0.57& -1.35&  0.52\\
& MSLE &  1.56  &0.98& -1.34&  0.52& -1.23  &0.48\\
\bottomrule
\end{tabular*}
\label{table:auc2}
\end{table}

\newpage

\begin{table}[!http]
\caption{RB ($\%$) and MSE ($\times 1000$) of eight methods for estimating the Youden index}
\centering
\small
\begin{tabular*}{\textwidth}{c@{\extracolsep{\fill}} c c c c c c c} 
\toprule
 &$(n_0,n_1)$ & \multicolumn{2}{c}{$(50,50)$} &  \multicolumn{2}{c}{$(100,100)$} &  \multicolumn{2}{c}{$(150,50)$} \\ 
 \cline{3-8}
Distribution ($J$)&Method &RB & MSE &RB & MSE&RB & MSE \\
\hline
Beta (0.3) &BP&4.59&6.07&3.20&2.96&3.36&3.54\\
&Box-Cox & 7.35 & 7.10 & 5.75 &  3.48 & 3.08 & 4.65\\
&ZL & 5.76 &6.22 &3.69 &2.96& 4.36 &3.63\\
&LZL&23.60&11.63&15.74&5.58&19.37&7.37\\
& ECDF & 23.61 & 11.64 & 15.76 & 5.58 & 19.39 & 7.38\\
& MNLE  & 23.61 & 11.64 & 15.76 & 5.58 & 19.39 & 7.38\\
& Kernel & 0.51 & 5.98 & 0.46 & 3.04 & 0.00 & 3.72\\
& MSLE & 0.51 & 5.98 & 0.46 & 3.04 & 0.00 & 3.72\\
\midrule
Beta (0.5) &BP&2.35&5.23&1.28&2.59&1.08&2.76\\
&Box-Cox & 8.64 &7.60 &7.34& 4.21& 8.66& 5.13\\
&ZL & 3.86& 5.55 &2.16& 2.64& 2.06 &2.89\\
&LZL&11.26&8.78&7.24&4.29&8.78&5.14\\
& ECDF & 11.27 & 8.78  &7.25&  4.30 & 8.79&  5.15\\
& MNLE &11.27  &8.78  &7.25 & 4.30 & 8.79  &5.15\\
& Kernel & -1.29&  5.30& -1.69&  2.76 &-1.71 & 3.17\\
& MSLE &  -1.29  &5.30& -1.69 & 2.76& -1.71  &3.17\\
\midrule
Beta (0.7) &BP&1.31&3.67&0.61&1.72&0.51&1.80\\
&Box-Cox & 8.20& 6.27& 7.60& 4.28& 7.84& 4.65\\
&ZL & 2.71& 4.07 &1.44 &1.86& 1.89 &4.71\\
&LZL&6.13&5.84&4.08&2.78&5.03&3.42\\
& ECDF &6.14 &5.84 &4.08& 2.79&5.03 &3.43\\
& MNLE & 6.14 &5.84& 4.08& 2.79& 5.03& 3.43\\
& Kernel & -1.39 & 3.71 &-1.47  &1.88& -1.20&  2.19\\
& MSLE & -1.39 & 3.71 &-1.47  &1.88 &-1.20  &2.19\\
\bottomrule
\end{tabular*}
\label{table:youden2}
\end{table}

\newpage

\begin{table}[!http]
\caption{RB ($\%$) and MSE ($\times 1000$) of eight methods for estimating the optimal cutoff point}
\centering
\small
\begin{tabular*}{\textwidth}{c@{\extracolsep{\fill}} c c c c c c c} 
\toprule
 &$(n_0,n_1)$ & \multicolumn{2}{c}{$(50,50)$} &  \multicolumn{2}{c}{$(100,100)$} &  \multicolumn{2}{c}{$(150,50)$} \\ 
 \cline{3-8}
Distribution ($J$)&Method &RB & MSE &RB & MSE&RB & MSE \\
\midrule
Beta (0.3) &BP&2.18&0.95&2.03&0.61&2.37&0.79\\
&Box-Cox & -0.41 & 1.33 & -0.30 &  0.67 & -2.03 & 0.88\\
&ZL & 0.24 &2.82& 0.71 &1.42& 0.89 &1.81\\
&LZL&0.97&7.73&1.06&5.59&1.86&6.66\\
& ECDF & 0.97 & 7.30 & 1.05 & 5.25 & 1.86 & 6.54\\
& MNLE & 1.10 & 7.21 & 1.08 & 5.22 & 1.87 & 6.53 \\
& Kernel & -0.74 & 5.07 & 0.44 & 3.10 & -0.14 & 4.09\\
& MSLE &-0.74&5.08&-0.45&3.10&-0.14&4.09\\
\midrule
Beta (0.5) &BP&1.13&0.62&0.96&0.33&1.10&0.49\\
&Box-Cox & -1.66&  0.72& -1.78&  0.42 &-2.97 & 0.75\\
&ZL & 0.08 &1.42 &0.36 &0.68 &0.73& 0.83\\
&LZL&0.15&3.84&0.10&2.46&0.94&3.03\\
& ECDF & 0.06 &3.68 &0.02 &2.40& 0.96 &2.99\\
& MNLE & 0.15& 3.68 &0.04 &2.38& 0.96&2.99\\
& Kernel & -1.24  &1.88& -1.07 & 1.13 &-0.89&  1.56\\
& MSLE &-1.24&1.88&-1.07&1.14&-0.90&1.56\\
\midrule
Beta (0.7) &BP&0.46&0.44&0.44&0.24&0.62&0.34\\
&Box-Cox & -1.65&  0.53 &-1.77  &0.36 &-2.02&  0.55\\
&ZL & -0.24 & 0.91&  0.06  &0.45 & 0.34 & 0.63\\
&LZL&-0.20&2.12&0.01&1.32&0.80&1.50\\
& ECDF & -0.46 & 1.93& -0.12&  1.26&  0.68 & 1.47\\
& MNLE & -0.39 & 1.93 &-0.09&  1.25&  0.69&  1.47\\
& Kernel & -1.16  &0.99& -0.82&  0.60 &-0.78 & 0.84\\
& MSLE &-1.17&1.00&-0.82&0.60&-0.78&0.84\\
\bottomrule
\end{tabular*}
\label{table:cutoff2}
\end{table}

\newpage
\section{Proof of Proposition 1 in the main article} \label{supp-section-2}

For (a). The proof is straightforward and we omit the details. 

For (b). Let $l_2(\alpha_0,\ldots,\alpha_N)=\log L_2(\alpha_0,\ldots,\alpha_N)$. Since there is no constraint on $\alpha_0$, and $l_2(\alpha_0,\ldots,\alpha_N)$ is a concave function of $(\alpha_0,\ldots,\alpha_N)$, $\hat\alpha_0$ has to satisfy the first-order condition. That is
\begin{equation}
\label{l2.deriv}
\frac{\partial l_2(\hat \alpha_0,\hat\alpha_1,\ldots,\hat\alpha_N)}{\partial\alpha_0}=0. 
\end{equation}
Note that 
$$
\frac{\partial }{\partial\alpha_0} l_2(\alpha_0,\alpha_1,\ldots,\alpha_N)
=\sum_{i=1}^mb_i-\sum_{i=1}^m(a_i+b_i)\theta(t_i). 
$$
Therefore, \eqref{l2.deriv} leads to 
$$
0=\sum_{i=1}^mb_i-\sum_{i=1}^m(a_i+b_i)\hat\theta(t_i), 
$$
which, together with the facts that $\sum_{i=1}^mb_i=n_1$ and  $\lambda=n_1/n$, 
imply that $\sum_{i=1}^m\hat\phi_i \hat\theta(t_i)=\lambda $. 
This finishes the proof. 

\section{Proof of Theorem 1 in the main article} \label{supp-section-3}

We first define some notation. We denote
\begin{eqnarray*}
d^2(\hat\theta, \theta_0)=
\int_{0}^1 \left\{\hat\theta(x)- \theta_0(x)\right\}^2~dG(x). 
\end{eqnarray*}
Let  $R(x)=\log\{f_1(x)/f_0(x)\}$, 
\begin{equation}
\label{def.thetaN}
\theta_N(x)=\frac{\lambda \exp\left \{ \sum_{l=0}^N R(l/N) B_l(x;N)\right \}}{1-\lambda+\lambda \exp\left \{ \sum_{l=0}^N R(l/N) B_l(x;N)\right \}},
\end{equation} 
and 
$$\mathcal{B}_N=\left\{
\theta(x)=
\frac{\lambda \exp\left\{ \sum_{l=0}^N \beta_lB_l(x;N) \right\} }
{1-\lambda+\lambda \exp\left\{ \sum_{l=0}^N \beta_lB_l(x;N)\right\} }: \beta_0\leq \beta_1\dots\leq \beta_N\right\}.$$
By Condition A3, $\theta_N(x)\in \mathcal{B}_N$. 
Further let 
$$
\ell_n(\theta)= \frac{1}{n}\sum_{i=1}^m \log [ \{ \theta(t_i) \}^{b_i}\{ 1-\theta(t_i)\} ^{a_i}].
$$
Then 
$$
\hat\theta(x)=\arg\max_{\theta(x)\in \mathcal{B}_N} \ell_n(\theta).
$$

By Condition A1, $n_1/n\to\lambda\in(0,1)$ as $n\to\infty$. 
As we discussed in Section 2.1 of the main paper, we write $\lambda = n_1/n$ and assume that it is constant, since it does not affect our technical development. 
Define 
$$\gamma_0(x;\theta)=4(1-\lambda)\left\{\sqrt{\frac{1-\theta(x)}{1-\theta_0(x)}}-1\right\}$$
and 
$$\gamma_1(x;\theta)=4\lambda\left\{\sqrt{\frac{\theta(x)}{\theta_0(x)}}-1\right\}.$$

The proof of Theorem 1 consists of three steps.
\begin{itemize}
\item[]{\bf Step 1.} We argue that 
\begin{eqnarray*}\label{step1}
   d^2(\hat\theta, \theta_0)
  \leq \int\gamma_0(x;\hat\theta)~d\{\tilde F_{0}(x)-F_{0}(x)\}+\int \gamma_1(x;\hat\theta)~d\{\tilde F_{1}(x)-F_{1}(x)\}+2\{ \ell_n(\theta_0)-\ell_n(\theta_N)\},
\end{eqnarray*}
where $\tilde F_0(x)$ and $\tilde F_1(x)$
are the empirical cumulative distribution functions of $\{X_1,\ldots,X_{n_0}\}$  and $\{Y_1,\ldots,Y_{n_1}\}$, respectively. 
\item[]{\bf Step 2.} We use results from empirical processes to show that 
\begin{equation}
\label{d2.part1}
\int\gamma_0(x;\hat\theta)~d\{\tilde F_{0}(x)-dF_{0}(x)\}=o_p(1)
\end{equation}
and 
\begin{equation}
\label{d2.part2}
\int \gamma_1(x;\hat\theta)~d\{\tilde F_{1}(x)-dF_{1}(x)\}=o_p(1).
\end{equation}
\item[]{\bf Step 3.}
We further show that  
\begin{equation}
\label{d2.part3}
 \ell_n(\theta_0)-\ell_n(\theta_N)=o_p(1). 
\end{equation}

\end{itemize}

We start with Step 1. 
Since $\hat\theta(x)$ maximizes $\ell_n(\theta)$ over $\mathcal{B}_N$ and $\theta_N(x)\in \mathcal{B}_N$, we have 
\begin{eqnarray*}
     \ell_n(\theta_0) -\ell_n(\theta_N)&\geq& \ell_n(\theta_0) -\ell_n(\hat\theta)\\
     &=& -\frac{1}{n}\sum_{i=1}^m\left\{b_i\log \frac{\hat\theta(t_i)}{\theta_0(t_i)}+a_i\log\frac{1-\hat\theta (t_i )}{1-\theta_0(t_i)}\right\}\\
    &=& -\frac{1}{n}\left\{\sum_{i=1}^{n_1}\log \frac{\hat\theta(Y_i)}{\theta_0(Y_i)}+\sum_{j=1}^{n_0}\log\frac{1-\hat\theta (X_j)}{1-\theta_0(X_j)}\right\}\\
    &=& -\left\{\lambda\int\log \frac{\hat\theta (x)}{\theta_0(x)}~d\tilde F_{1}(x)+(1-\lambda)\int\log\frac{1-\hat\theta(x)}{1-\theta_0(x)}~d\tilde F_{0}(x)\right\}.
\end{eqnarray*}

Applying the inequality $0.5\log x\leq \sqrt{x}-1$ for $x>0$, we further get
\begin{eqnarray*}
    2\{\ell_n(\theta_0) -\ell_n(\theta_N)\} &\geq& -\int \gamma_1(x;\hat\theta)~d\tilde F_{1}(x)-\int \gamma_0(x;\hat\theta)~d\tilde F_{0}(x)\\
    &=& -\int \gamma_1(x;\hat\theta)~d\{\tilde F_{1}(x)-dF_1(x)\}
    -\int \gamma_1(x;\hat\theta)~dF_1(x)\\
    &&-\int \gamma_0(x;\hat\theta)~d\{\tilde F_{0}(x)-F_0(x)\}
      -\int \gamma_0(x;\hat\theta)~dF_0(x),
\end{eqnarray*}
which leads to 
\begin{eqnarray*}
     -\int \gamma_0(x;\hat\theta)~dF_0(x)-\int \gamma_1(x;\hat\theta)~dF_1(x)
     &\leq& \int \gamma_1(x;\hat\theta)~d\{\tilde F_{1}(x)-dF_1(x)\}\\
     &&+ \int \gamma_0(x;\hat\theta)~d\{\tilde F_{0}(x)-F_0(x)\}\\
     &&+ 2\{\ell_n(\theta_0) -\ell_n(\theta_N)\}. 
\end{eqnarray*}
To finish the proof of Step 1, it suffices to show that
\begin{equation}
\label{d2.step0}
    d^2(\hat\theta, \theta_0)\leq -\int \gamma_0(x;\hat\theta)~dF_0(x)-\int \gamma_1(x;\hat\theta)~dF_1(x).
\end{equation}

By the definition of $\theta_0(x)$, we have that
\begin{align}\label{measure1}
    dF_1(x)=\frac{\theta_0(x)/\{ 1-\theta_0(x)\}}{\lambda/(1-\lambda)}dF_0(x)
\end{align}
and
\begin{align}\label{measure2}
    dG(x)=\lambda dF_1(x)+(1-\lambda) dF_0(x)=\frac{1-\lambda}{1-\theta_0(x)}dF_0(x).
\end{align}
Note that (\ref{measure1}) and (\ref{measure2}) imply
\begin{align*}
    -\int \gamma_1(x;\hat\theta)~dF_1(x)&=\int 4\lambda \frac{\theta_0(x)/\{1-\theta_0(x)\}}{\lambda/(1-\lambda)}\left\{1-\sqrt{\frac{\hat\theta(x)}{\theta_0(x)}}\right\}~dF_0(x)\\
    &=4\int \theta_0(x)\left\{1-\sqrt{\frac{\hat\theta(x)}{\theta_0(x)}}\right\}~dG(x)
\end{align*}
and
\begin{align*}
    -\int \gamma_0(x;\hat\theta)~dF_0(x)&=4\int \{ 1-\theta_0(x)\}\left\{1-\sqrt{\frac{1-\hat\theta(x)}{1-\theta_0(x)}}\right\}~dG(x).
\end{align*}
Hence,
\begin{align*}
    &-\int \gamma_0(x;\hat\theta)~dF_0(x)-\int \gamma_1(x;\hat\theta)~dF_1(x)\\
    = & 4\int\left[ 1-\sqrt{\hat\theta(x)\theta_0(x)}-\sqrt{\left\{1-\hat\theta(x)\right\}\left\{1-\theta_0(x)\right\}}\right]~dG(x)\\
    =&2\int\left\{\sqrt{\hat\theta(x)}-\sqrt{\theta_0(x)}\right\}^2~dG(x)
    +2\int\left\{\sqrt{1-\hat\theta(x)}-\sqrt{1-\theta_0(x)}\right\}^2~dG(x).
\end{align*}
Using the facts that both $\theta_0(x)$ and $\hat\theta(x)\in[0,1]$, 
we  get that 
\begin{align*}
    &-\int \gamma_0(x;\hat\theta)~dF_0(x)-\int \gamma_1(x;\hat\theta)~dF_1(x)\\
    \geq&2\int\frac{1}{4}\left\{\hat\theta(x)-\theta_0(x)\right\}^2~dG(x)+2\int\frac{1}{4}\left\{\hat\theta(x)-\theta_0(x)\right\}^2~dG(x)\\
    =&d^2(\hat\theta, \theta_0).
\end{align*}
This finishes the proof of \eqref{d2.step0}, and hence that of Step 1.

We now move to Step 2. 
Let 
$$\mathcal{H}=\{h(x): h(x)\text{ is non-decreasing on [0,1]}\mbox{ and } 
h(x)\in[0,1]\}.$$
Using Theorems 2.2 and  9.24 of \cite{kosorok2008introduction}, 
we have that the class $\mathcal{H}$ is $P$-Glivenko--Cantelli ($P$-G-C) for $P=F_0$ and $F_1$. That is, 
\begin{equation*}
    \sup_{h(x)\in\mathcal{H}} \left|\int h(x)~d\{\tilde F_{0}(x)-F_0(x)\}\right|=o_p(1)~\mbox{ and }~
    \sup_{h(x)\in\mathcal{H}}\left|\int h(x)~d\{\tilde F_{1}(x)-F_1(x)\}\right|=o_p(1).
\end{equation*}
Since $\mathcal{B}_N\subset \mathcal{H}$, the class $\mathcal{B}_N$ is also $P$-G-C for $P=F_0$ and $F_1$. 

Let 
$$
{\Gamma}_0
=\{\gamma_0(x;\theta): \theta(x)\in\mathcal{B}_N\}.$$
We notice that 
$|\gamma_0(x;\theta)|\leq 4(1-\lambda)[1/\{1-\theta_0(x)\}+1]$ and
\begin{align*}
\int 4(1-\lambda)[1/\{1-\theta_0(x)\}+1]~dF_0(x)=&4(1-\lambda)+\int 4(1-\lambda)/\{1-\theta_0(x)\}~dF_0(x)\\
    =&4(1-\lambda)+\int 4~dG(x)\\
    =&4+4(1-\lambda)<\infty.
\end{align*}
That is, the class ${\Gamma}_0$ has an integrable envelope. 
 Applying  Corollary 9.27 of \cite{kosorok2008introduction}, we then get that $\Gamma_0$ is $F_0$-G-C, i.e.,
 $$
\sup_{\theta(x)\in \Gamma_0}\left|\int \gamma_0(x;\theta) d\{\tilde F_0(x)-F_0(x)\}
\right|=o_p(1),
$$
which implies that 
 $$
\int \gamma_0(x;\hat\theta) d\{\tilde F_0(x)-F_0(x)\}
=o_p(1).
$$
Similarly, 
 $$
\int \gamma_1(x;\hat\theta) d\{\tilde F_1(x)-F_1(x)\}
=o_p(1).
$$
This finishes Step 2. 

We consider Step 3. 
It can be checked that 
\begin{eqnarray*}
 \ell_n(\theta_0)-\ell_n(\theta_N)&=&
 -\frac{\lambda}{n_1}\sum_{i=1}^{n_1}\log \frac{\theta_N(Y_i)}{\theta_0(Y_i)}
 -
 \frac{1-\lambda}{n_0}\sum_{j=1}^{n_0}\log\frac{1-\theta_N(X_j)}{1-\theta_0(X_j)}\\
 &=&-\lambda\int \log \frac{\theta_N(x)}{\theta_0(x)}d\tilde F_1(x)
 -(1-\lambda)\int \log \frac{1-\theta_N(x)}{1-\theta_0(x)}d\tilde F_0(x).
 \end{eqnarray*}
By Condition A3 and the construction of $\theta_N(x)$  
in \eqref{def.thetaN}, 
we have 
$$
\theta_0(x)\in[\delta,1-\delta]
~\mbox{ and }~
\theta_N(x)\in[\delta,1-\delta]
,
$$
where $\delta>0$ is provided in Condition A3. 

Let $$\mathcal{H}^*=\{h(x): h(x)\text{ is non-decreasing on [0,1]}\mbox{ and } 
h(x)\in[\delta,1-\delta]\}.$$
Then $\mathcal{H}^*\subset \mathcal{H}$.
Hence,  the class $\mathcal{H}^*$
is $P$-G-C for $P=F_0$ and $F_1$. 
Note that for any $h(x)\in \mathcal{H}^*$, 
both $|\log\{h(x)/\theta_0(x)\}|$ and 
$|\log[\{1-h(x)\}/\{1-\theta_0(x)\}]|$
are bounded functions. 
Using  Corollary 9.27 of \cite{kosorok2008introduction}, 
we further have 
the two classes
$$
\left\{
\log\frac{h(x)}{\theta_0(x)}
: h(x)\in \mathcal{H}^*
\right\}
~
\mbox{ and }
~
\left\{
\log\frac{1-h(x)}{1-\theta_0(x)}
: h(x)\in \mathcal{H}^*
\right\}
$$
are also $P$-G-C for $P=F_0$ and $F_1$, 
which implies that 
$$
\int \log \frac{\theta_N(x)}{\theta_0(x)}d\{\tilde F_1(x)-F_1(x)\}
=o_p(1)
~
\mbox{ and }
~
\int \log \frac{1-\theta_N(x)}{1-\theta_0(x)}d\{\tilde F_0(x)-F_0(x)\}=o_p(1).
$$
Hence, 
\begin{eqnarray*}
 \ell_n(\theta_0)-\ell_n(\theta_N)&=&
 -\lambda\int \log \frac{\theta_N(x)}{\theta_0(x)}d F_1(x)
 -(1-\lambda)\int \log \frac{1-\theta_N(x)}{1-\theta_0(x)}d  F_0(x)+o_p(1)\\
 &=&
 -
 \int\left[\theta_0(x) \log \frac{\theta_N(x)}{\theta_0(x)}
 +\{1-\theta_0(x)\}\log \frac{1-\theta_N(x)}{1-\theta_0(x)}
 \right]dG(x)+o_p(1).
 \end{eqnarray*}
To finish Step 3, it is sufficient to show that 
$$
 \int\left[\theta_0(x) \log \frac{\theta_N(x)}{\theta_0(x)}
 +\{1-\theta_0(x)\}\log \frac{1-\theta_N(x)}{1-\theta_0(x)}
 \right]dG(x)
 =o(1). 
$$

Recall that $R(x)=\log\{f_1(x)/f_0(x)\}$.
Let 
$$
R_N(x)= \sum_{l=0}^N R(l/N) B_l(x;N).
$$
According to \cite{lorentz2013bernstein},
under Condition A3, we have 
\begin{equation}
\label{diff.RRN}
\lim_{N\to\infty}\sup_{x\in [0,1]}\left| R(x)-R_N(x)\right|=0.
\end{equation}
Further let 
$$
g(t)=\frac{\lambda \exp(t)}{1-\lambda+\lambda \exp(t)}. 
$$
Then, $\theta_0(x)=g\big(R(x)\big)$ and $\theta_N(x)=g\big(R_N(x)\big)$. It can be easily verfied that $|g'(t)|\leq 1$ for all $t$, which, together with  \eqref{diff.RRN}, implies that 
\begin{equation}
\label{diff.ttN}
\lim_{N\to\infty}\sup_{x\in [0,1]}\left| \theta_N(x)-\theta_0(x)\right|=0.
\end{equation}
Using the second-order Taylor expansion, we get 
\begin{eqnarray}
\nonumber
&&
 \left|\int\left[\theta_0(x) \log \frac{\theta_N(x)}{\theta_0(x)}
 +\{1-\theta_0(x)\}\log \frac{1-\theta_N(x)}{1-\theta_0(x)}
 \right]dG(x)\right|  \\
 &=&
 0.5\int\left[
 \frac{\theta_0(x)}{\xi_N^2(x)}
 +
 \frac{1-\theta_0(x)}{\{1-\xi_N(x)\}^2}
 \right] \left\{
 \theta_N(x)-\theta_0(x)
 \right\}^2dG(x)\nonumber\\
 &\leq&\frac{0.5}{\delta^2}
 \int\left\{
 \theta_N(x)-\theta_0(x)
 \right\}^2
 dG(x)
 \label{d2.part3.upper}
\end{eqnarray}
where $\xi_N(x)$ is a function between $\theta_0(x)$ and $\theta_N(x)$, and we have used Condition A3 in the last step.
Combining \eqref{diff.ttN} and \eqref{d2.part3.upper}
leads to 
$$
 \left|\int\left[\theta_0(x) \log \frac{\theta_N(x)}{\theta_0(x)}
 +\{1-\theta_0(x)\}\log \frac{1-\theta_N(x)}{1-\theta_0(x)}
 \right]dG(x)\right| 
 \leq \frac{0.5}{\delta^2} 
\left\{ 
\sup_{x\in [0,1]}\left| \theta_N(x)-\theta_0(x)\right|
\right\}^2
\to 0
$$
as $N\to\infty$.
This finishes the proof of Step 3, and hence the proof of Theorem 1. 

\section{Proof of Theorem 2 in the main article} \label{supp-section-4}

\textbf{Part (a)} 
We concentrate on the consistency of $\hat F_1(x)$; 
the consistency of $\hat F_0(x)$ follows similarly.

Let 
$$
\tilde G(x)=
\lambda \tilde F_1(x)+(1-\lambda)\tilde F_0(x)
=
\sum_{i=1}^m \frac{a_i+b_i}{n}I(t_i\leq x).
$$
Note that for any $u$, 
\begin{align*}
    \hat F_1(u)&=\sum_{i=1}^m \hat p_{i1}I(t_i\leq u)\\
    &=\sum_{i=1}^m \frac{1}{\lambda}\frac{a_i+b_i}{n}\hat\theta(t_i)I(t_i\leq u)\\
    &= \int \frac{1}{\lambda} \hat\theta(x)I(x\leq u)~d\tilde G(x).
\end{align*}
Recall 
\begin{align*}
    F_1(u)&=\int \frac{1}{\lambda} \theta_0(x)I(x\leq u)~dG(x).
\end{align*}
Then 
\begin{align}\label{term1}
    \hat F_1(u)-F_1(u)&=\frac{1}{\lambda}\int  \{ \hat\theta(x)-\theta_0(x)\}I(x\leq u)~dG(x)\\\label{term2}
    &+\frac{1}{\lambda}\int  \hat\theta(x)I(x\leq u)~d\{\tilde G(x)-G(x)\}. 
\end{align}
In the following, 
we show that the two terms in (\ref{term1}) and (\ref{term2}) are $o_p(1)$. 

We first consider (\ref{term1}). 
Note that 
\begin{align}
\nonumber
   \left |\frac{1}{\lambda}\int  \{ \hat\theta(x)-\theta_0(x)\}I(x\leq u)~dG(x)\right |
    \leq &\frac{1}{\lambda} \int  |\hat\theta(x)-\theta_0(x)|I(x\leq u)~dG(x)\\\nonumber
    \leq &\frac{1}{\lambda} \int  |\hat\theta (x)-\theta_0(x)|~dG(x)\\\nonumber
    \leq &\frac{1}{\lambda} \left[\int  \left\{ \hat\theta(x)-\theta_0(x)\right\}^2~dG(x)\right]^{1/2}\\
    = & 
    \lambda^{-1}
    d(\hat\theta,\theta_0)=o_p(1)\label{res1},
\end{align}
where we have used the result in Theorem 1.

Next, we consider(\ref{term2}).
According to Lemmas 9.8 and 9.12 of \cite{kosorok2008introduction},
$$\left\{I\big(x\leq u\big):~u\in \mathcal{R}\right\}$$
 is a Vapnik--Chervonenkis subgraph class
with the bounded envelope  1.
By Theorems 8.14 and 9.2 of \cite{kosorok2008introduction}, we have that $\{I(x\leq u):~u\in \mathcal{R}\}$ is $P$-G-C for both $F_0$ and $F_1$.
Recall that in the proof of Theorem 1, we have argued that  $\mathcal{B}_N$ is also $P$-G-C for both $F_0$ and $F_1$.
Then by Corollary 9.27 of \cite{kosorok2008introduction}, $\{I(x\leq u)\theta(x):~\theta\in \mathcal{B}_N,~u\in R\}$ is $P$-G-C  for both $F_0$ and $F_1$.
That is,
\begin{equation}\label{GCF0}
    \sup_{\theta(x)\in\mathcal{B}_N,~u\in\mathcal{R}}
    \left|\int I(x\leq u)\theta(x)~d\{\tilde F_{0}(x)-F_0(x)\}\right|=o_p(1)
\end{equation}
and
\begin{equation}\label{GCF1}
    \sup_{\theta(x)\in\mathcal{B}_N,~u\in\mathcal{R}}
    \left|\int I(x\leq u)\theta(x)~d\{\tilde F_{1}(x)-F_1(x)\}
    \right|=o_p(1).
\end{equation}
Note that  (\ref{GCF0}) and (\ref{GCF1}) together lead to 
\begin{equation*}
    \sup_{\theta(x)\in\mathcal{B}_N,~u\in\mathcal{R}}
    \left|\int I(x\leq u)\theta(x)~d\{\tilde G(x)-G(x)\}\right|=o_p(1),
\end{equation*}
which implies that  
\begin{equation}\label{GCF}
    \sup_{u\in[0,1]}
    \left|\int I(x\leq u)\hat\theta(x)~d\{\tilde G(x)-G(x)\}\right|=o_p(1).
\end{equation}
Combining the results in (\ref{term1})--(\ref{res1}) and (\ref{GCF}), we have 
\begin{equation*}
    \sup_{x\in [0,1]} |\hat F_1(x)-F_1(x)|=o_p(1).
\end{equation*}
This completes the proof of Part (a). 

\textbf{Part (b)} The proof for this part is very similar to the proof of Theorem 1 in \cite{jokiel2013nonparametric}. We omit the details here.

\textbf{Part (c)} Note that
\begin{align*}
    |\widehat{AUC}-AUC|&=\left|\int_0^1\widehat{ROC}(s)-ROC(s)~ds\right|\\
    &\leq \int_0^1\left|\widehat{ROC}(s)-ROC(s)\right|ds\\
    &\leq \int_0^1\sup_{s\in[0,1]}\left|\widehat{ROC}(s)-ROC(s)\right|ds\\
    &\leq\sup_{s\in[0,1]}\left|\widehat{ROC}(s)-ROC(s)\right|\\
    &=o_p(1).
\end{align*}

\textbf{Part (d)}
Let $A_n=\left\{\inf_{x\in[0,1]}\hat\theta(x)<\lambda<\sup_{x\in[0,1]}\hat\theta(x)\right\}$. 
We first show that 
\begin{equation}
\label{limit.An}
\lim_{n\to\infty}P(A_n)=1,
\end{equation}
which implies that the probability that the solution of $\hat\theta(x)=\lambda$ exists in $[0,1]$ approaches to $1$ as $n$ goes to infinity.
Define
$$A_{n1}=\left\{\inf_{x\in[0,1]}\hat\theta(x)\geq\lambda\right\}$$
and
$$A_{n2}=\left\{\sup_{x\in[0,1]}\hat\theta(x)\leq\lambda\right\}.$$
It suffices to show that
\begin{align}
    \label{inf}\lim_{n\to\infty}P(A_{n1})&=0,
\end{align}
and
\begin{align}
    \label{sup}\lim_{n\to\infty}P(A_{n2})&=0.
\end{align}
We concentrate on  \eqref{inf}; the proof for \eqref{sup}follows similarly.

By Condition A4, there exists $\epsilon_0>0$ such that $\theta_0(x)$ is strictly increasing on $[C-\epsilon_0,C+\epsilon_0]$.
Hence, $\theta_0(C)-\theta_0(C-\epsilon_0)>0$.
By the continuity and monotonicity of $\theta_0(x)$, there exists $0<\delta_0<\epsilon_0$ 
such that 
$$0<\theta_0(x)-\theta_0(C-\epsilon_0)\leq\frac{\theta_0(C)-\theta_0(C-\epsilon_0)}{2}$$
for any $x\in[C-\epsilon_0,C-\epsilon_0+\delta_0]$,
which together with the fact  $\theta_0(C)=\lambda$  imply that 
\begin{equation}
\label{An1.lower}
\lambda-\theta_0(x)=\theta_0(C)-\theta_0(x)\geq\frac{\theta_0(C)-\theta_0(C-\epsilon_0)}{2},~\forall x\in[C-\epsilon_0,C-\epsilon_0+\delta_0].
\end{equation}

Let $\delta_1=0.25{\delta_0\left\{\theta_0(C)-\theta_0(C-\epsilon_0)\right\}^2}>0$.
Then \eqref{An1.lower} implies that 
\begin{align*}
    P(A_{n1})\leq&P\left(\int_{C-\epsilon_0}^{C-\epsilon_0+\delta_0}\left\{
    \hat\theta(x)-\theta_0(x)\right\}^2~dx\geq \delta_1\right)\\
    \leq&P\left(\int_{0}^{1}\left\{
    \hat\theta(x)-\theta_0(x)\right\}^2~dx\geq \delta_1\right)\\
    =&P\left(d^2(\hat\theta,\theta_0)\geq \delta_1\right).
\end{align*}
By Theorem 1, 
we get 
\begin{align*}
\lim_{n\to\infty}P(A_{n1})=0.
\end{align*}
This finishes the proof of \eqref{inf} and hence that of \eqref{limit.An}.
With the result in \eqref{limit.An},
 without loss of generality, we assume that the solution to 
 $\hat\theta(x)=\lambda$ exists, i.e., 
$$
\hat\theta(\hat C)=\lambda.
$$

We now move to the consistency of $\hat C$. 
Equivalently, we need to show that for any $\epsilon>0$,  
\begin{align}\label{large}
    \lim_{n\to \infty}P(\hat C\geq C+\epsilon)=0,
\end{align}
and
\begin{align}\label{less}
    \lim_{n\to \infty}P(\hat C\leq C-\epsilon)=0.
\end{align}

We first consider (\ref{large}).
Note that 
\begin{align*}
    P(\hat C\geq C+\epsilon)\leq&P\left(\hat\theta(\hat C)\geq \hat\theta(C+\epsilon)\right)\\
    =&P\left( \hat\theta(C+\epsilon)-\lambda\leq 0\right)\\
    \leq&P\left(\int_{C}^{C+\epsilon}\left\{
    \theta_0(x)-\hat\theta(x)\right\}^2~dx
    \geq \int_{C}^{C+\epsilon}\left\{
    \theta_0(x)-\lambda\right\}^2~dx\right)\\
    \leq&P\left(\int_{0}^{1}\left\{\theta_0(x)-\hat\theta(x)\right\}^2~dx\geq \int_{C}^{C+\epsilon}\left\{
    \theta_0(x)-\lambda\right\}^2~dx \right)
    \\
    =&P\left(d^2(\hat\theta,\theta_0)\geq \int_{C}^{C+\epsilon}\left\{
    \theta_0(x)-\lambda\right\}^2~dx \right). 
\end{align*}
Condition A4 implies that  $\int_{C}^{C+\epsilon}\left\{
    \theta_0(x)-\lambda\right\}^2~dx>0 $. 
   By Theorem 1, 
   we have 
\begin{align*}
    \lim_{n\to\infty}P(\hat C\geq C+\epsilon)=0.
\end{align*}
This completes \eqref{large}. The proof of \eqref{less} follows similarly. 

Finally, we consider the consistency of $\hat J$. 
Note that 
\begin{align*}
    \hat J-J=&\left\{\hat F_0(\hat C)-\hat F_1(\hat C)\right\}-\left\{F_0(C)-F_1(C)\right\}\\
    =&\left\{\hat F_0(\hat C)-\hat F_1(\hat C)\right\}-\left\{F_0(\hat C)-F_1(\hat C)\right\}\\
    &+\left\{F_0(\hat C)-F_1(\hat C)\right\}-\left\{F_0(C)-F_1(C)\right\}.
    \end{align*}
Then 
\begin{align}
|\hat J-J|\leq & \left|F_0(\hat C)-F_0(C)\right|+\left| F_1(\hat C)-F_1(C)\right|\label{hatJ.part1}\\
    &+\sup_{x\in[0,1]}\left|\hat F_0(x)-F_0(x)\right|+\sup_{x\in[0,1]}\left|\hat F_1(x)-F_1(x)\right|\label{hatJ.part2}. 
\end{align}
By the consistency of $\hat C$ and the continuous mapping theorem, 
the term in  (\ref{hatJ.part1}) is $o_p(1)$. 
The term in (\ref{hatJ.part2}) is also $o_p(1)$ by the results in Part (a).
Hence $\hat J-J=o_p(1)$  as claimed.

\end{document}